\documentclass[10pt]{iopart}
\usepackage{mathrsfs}
\usepackage{iopams}
\usepackage{cite}
\usepackage{epsfig}
\usepackage{graphicx}
\usepackage{color}
\def\cqg{{\em Class. Quantum Grav.\/} }
\def\grg{{\em Gen. Rel. Grav.\/} }
\def\prd{{\em Phys. Rev.\/} D }

\def\pre{{\em Phys. Reports\/} }
\def\apj{{\em Astrophys. J.\/} }
\def\jmp{{\em J. Math. Phys.\/} }
\def\mn{{\em Mon. Not. Roy. Astr. Soc.\/} }
\def\assb{{\em Ann. Soc. Sci. Bruxelles\/} }
\def\pnasu{{\em Proc. Nat. Acad. Sci. U.S.A.\/} }

\def\qjr{{\em Quart. J. R. Astron. Soc.\/} }

\def\aa{{\em Astron. Astrophys.\/} }

\def\qm{{\em Quaestiones Mathematicae\/} }

\begin{document}

\title{Obtaining the spacetime metric from cosmological
observations}

\author[H.-C. Lu {\it et al.}]{Teresa Hui-Ching Lu and Charles Hellaby} \vspace{3mm}

\address{Department of Mathematics and Applied Mathematics,
University of Cape Town, Rondebosch 7701, South Africa}

\date{\today}

\eads{Teresa.HuiChingLu@gmail.com and Charles.Hellaby@uct.ac.za}

\begin{abstract}
Recent galaxy redshift surveys have brought in a large amount of accurate cosmological data
out to redshift 0.3, and future surveys are expected to achieve a high degree of
completeness out to a redshift exceeding 1.  Consequently, a numerical programme for
determining the metric of the universe from observational data will soon become practical;
and thereby realise the ultimate application of Einstein's equations.
Apart from detailing the cosmic geometry, this would allow us to verify and quantify
homogeneity, rather than assuming it, as has been necessary up to now, and to do that on
a metric level, and not merely at the mass distribution level.
This paper is the beginning of a project aimed at such a numerical implementation.
The primary observational data from our past light cone consists of galaxy redshifts,
apparent luminosities, angular diameters and number densities, together with source
evolution functions, absolute luminosities, true diameters and masses of sources.
Here we start with the simplest case, that of spherical symmetry and a dust equation
of state, and execute an algorithm that determines the unknown metric functions from this data.
We discuss the challenges of turning the theoretical algorithm into a workable numerical
procedure, particularly addressing the origin and the maximum in the area distance.  Our
numerical method is tested with several artificial data sets for homogeneous and
inhomogeneous models, successfully reproducing the original models.
This demonstrates the basic viability of such a scheme.  Although current surveys don't
have sufficient completeness or accuracy, we expect this situation to change in the near
future, and in the meantime there are many refinements and generalisations to be added.
\end{abstract}

\pacs{98.80.-k, 98.65.-r}


\section{Introduction}

In modern cosmology, many attempts have been made to determine the
large-scale structure of the physical universe using constraints
provided by cosmological observations and knowledge derived from
local physical experiments. The most common approach is to adopt the
postulate that the universe is spatially homogeneous on large scales
--- the Friedmann-Lema\^{\i}tre-Robertson-Walker (FLRW) model.
Hence using observational data to determine the few free parameters
characteristic of such universe models has become the primary
objective, and this overall framework has been presented in great
detail in the literature.

The cosmological principle (see \cite{Bondi} and \cite{Weinberg72})
expresses spatial homogeneity as a point of principle, whereas the
copernican principle merely states that we are not privileged
observers.  As pointed out in \cite{Ellis75}, the cosmological
principle determines a complete universe model, but we cannot verify
it fully due to the predictions it makes about parts of the universe
far beyond our observations.  Although the copernican principle only
has implications for the observable universe, its validity can
potentially be proven with observations.  Despite this, it is the
much stronger cosmological principle which is almost invariably
assumed in practice. Certainly, a good argument for homogeneity is
provided by the Ehler-Geren-Sachs (EGS) theorem \cite{EGS68}, which
required exactly isotropic CMBR observations for all observers, and
the `almost EGS theorem', or Stoeger-Maartens-Ellis (SME) theorem
\cite{SME95EGS}, which allows small anisotropies in the cosmic
microwave background radiation (CMBR) and obtains an `almost FLRW'
geometry \cite{MES95}.  But our attempts to verify homogeneity
should not stop there.

The general assumption that the universe has a Robertson-Walker
metric on the very large scale has served cosmology well, and is
implicit in many calculations.  But this makes verifying homogeneity
rather tricky as there is a distinct danger of a circular argument.
To analyse the cosmological data consistently requires the use of a
non-homogeneous metric.

Current and planned galaxy surveys are vastly increasing the amount
of cosmological data available for analysis. Already in recent years
there has been a dramatic improvement in the number of measured
cosmological parameters and the accuracy of their values. Properties
of the matter distribution have been well studied, but always with
the assumption of a homogeneous background metric. As accurate
cosmological data accumulates, the proper reduction and
interpretation of the high redshift data will require knowledge of
the cosmic geometry that is traversed by the light rays we observe.
It will no longer be necessary to assume homogeneity, the data will
make it possible to quantify the level of homogeneity on different
scales. Hence, being able to {\em prove} the homogeneity of the
observable region of the universe rather than assuming it in
principle is a long term objective of the current project.  There
are of course a variety of methods for checking homogeneity, such as
the Sunyaev-Zel'dovich effect, and it is important to pursue the
full range of methods.

It should be emphasised that radial homogeneity is far harder to
prove clearly than isotropy. Our cosmological observations are
restricted to our past null cone, which is a 3-dimensional slice
through our 4-dimensional spacetime, and the expected variation of
observations with redshift is affected by the cosmic equation of
state, the evolution of the observed sources, and the geometry of
spacetime. Disentangling these effects without assuming homogeneity
is not a trivial exercise \cite{MHE,Hellaby01}.

We wish to determine the spacetime geometry as far as possible from
astronomical observations with minimal a priori assumptions. In
principle a set of observations of the redshifts, angular diameters,
and apparent luminosities of galaxies, as well as their number
counts, combined with knowledge of the cosmic equation of state and
the true diameters, luminosities and masses of the sources (and the
evolution of these source properties), can be turned into metric
information. The idea of reducing observed cosmological data to a
metric was first explicitly discussed by Kristian and Sachs
\cite{KristianS}; they examined how this could be done near our
present spacetime position by deriving expressions in power series
for some astronomical observations near the observer in a general
metric, and demonstrated the difficulties faced in confirming
homogeneity of the universe from observations. However, the problem
of source evolution was barely addressed in their derivations. In
the ideal observational cosmology program by Ellis and Stoeger and
others
\cite{ENMSW,SNME92,SEN92,SNE92b,SNE92c,MaartensMatravers,MHMS96,AS99,AABFS01,ARS01,RS03},
they took a slightly different approach to Kristian and Sachs as
they aimed to determine what could and could not be decidable in
cosmology on the basis of ideal astronomical observations, and so
considered the limits of verification in cosmology. They worked with
observational coordinates since all observational data are given,
not on the usual spacelike surface of constant time, but rather on
our past null cone, which is centred at our observational position
on our worldline. Hence, the observational data can be used with
ease in the implementation of any algorithm developed through using
the Einstein field equations (EFEs) that are written in
observational coordinates.

Thus there has been a fair bit of theoretical work on how to
determine the cosmic metric from standard obervations, but
implementation has not been attempted and the two key issues of
choosing appropriate numerical methods and handling real
observational data have not been properly addressed. However Bishop
and Haines \cite{BH96} did make a numerical attack on the problem
that was only partly successful. They treated the past null cone
(PNC) as a time-reversed characteristic initial value problem
(CIVP). Since the CIVP code is not intended to deal with a
reconverging past null cone, their numerics blew up at the maximum
in the diameter distance, and they were not able to extend past this
point%
\footnote{Their initial data on the PNC were the diameter distance,
the 2-metric derived from image distortions, the matter 4-velocity
derived from redshifts and proper motions, and the matter density
derived from number counts.  Knowledge of the true shapes, absolute
luminosities and masses of the sources was assumed. They adapted an
axially symmetric, zero pressure, zero rotation, zero $\Lambda$,
CIVP code and tested it on spherically symmetric Einstein-de Sitter
PNC data, finding it to be second order accurate. Although $r$ and
$l$ are called luminosity distances throughout, it is evident from
their equations and figure 4 that they are diameter distances. This
work has not been followed up.}.
As shown here, careful consideration of the nature of this maximum
allows the integration to be continued to much higher redshifts.
This is clearly a very big task, and will take years to develop into
a rigorous algorithm generating believable results.  We here
describe the beginnings of such a procedure, necessarily simple at
first.  Our focus here is on turning the theoretical algorithm
outlined in \cite{MHE} into a workable numerical method, and thereby
providing a demonstration of the viability of a key component of the
problem. In tackling a problem of this magnitude, it is essential to
start simply, which is the main reason for initially assuming
spherical symmetry and zero cosmological constant. In the long term
we envisage a much more general treatment. (We note however, that we
are unavoidably at the centre of our past null cone, and spherical
coordinates provide a natural description.) Even this simple first
step provides some interesting challenges, which are discussed
below. Our expectation is that the accuracy and especially the
completeness of cosmological redshift surveys will be much enhanced
in the coming years, and extracting the geometric implications of
the observations will become possible.

\section{The Lema\^{\i}tre-Tolman-Bondi Model and its null cone relations}

The general spherically symmetric metric for an irrotational dust
matter source in synchronous comoving coordinates is the
Lema\^{\i}tre-Tolman-Bondi (LTB) \cite{L,T,B} metric
\begin{equation}
ds^{2} = -dt^{2} + \frac{\ [R'(t,r)]^{2}\ }{1 + 2E(r)}\ dr^{2} +
R^{2}(t,r)d \Omega^{2}\ , \label{LTBmetric}
\end{equation}
where $R'(t,r) = \partial R(t,r)/\partial r$, and $d \Omega^2 = d
\theta^2 + \sin^2 \theta d \phi^2$.  The function $R
= R(t,r)$ is the areal radius, since the proper area of a sphere of
coordinate radius $ r $ on a time slice of constant $t$ is $4\pi
R^{2}$. The function $E = E(r) \geq -1/2$ is an arbitrary function
of the LTB model representing the local geometry.

Solving the EFEs with $\Lambda = 0$ gives us a generalised Friedmann
equation for $R(t,r)$,
\begin{equation}
\dot R^{2}(t,r) = \frac{2M(r)}{\ R(t,r)\ }+2E(r)\ , \label{Rdot}
\end{equation}
and an expression for the density
\begin{equation}
4\pi \rho(t,r) = \frac{M^{'}(r)}{\ R^{2}(t,r)R^{'}(t,r)\ }\ , \label{density}
\end{equation}
where $M(r)$ is another arbitrary function of the LTB
model that gives the gravitational mass within comoving radius $r$.
Here $E(r)$ also plays a dynamical role, it determines the local
energy per unit mass of the dust particles. Equation (\ref{Rdot})
can be solved in terms of a parameter $\eta = \eta(t,r)$, and a
third arbitrary function $t_B(r)$ which is the time of the big bang
locally:
\begin{equation}
\hspace{-15 mm} R = \frac{M}{\ 2E\ }\ (\cosh \eta - 1)\ ,\ \ \ \ \
\sinh \eta - \eta = \frac{\ (2E)^{3/2}(t - t_B)\ }{M}\ ;\ \ \ \ \ E
> 0\ ,\label{hyperbolic}
\end{equation}
\begin{equation}
\hspace{-15 mm} R = M \left( \frac{\eta^2}{2} \right)\ ,\ \ \
\ \ \left( \frac{\eta^3}{6} \right) = \frac{\ (t - t_B)\ }{M}\ ;\ \ \ \ \ E
= 0,\label{parabolic}
\end{equation}
\begin{equation}
\hspace{-15 mm} R = \frac{M}{\ (-2E)\ }\ (1 - \cos \eta)\ ,\ \ \ \ \
\eta - \sin \eta = \frac{\ (-2E)^{3/2}(t - t_B)\ }{M}\ ;\ \ \ \ \ E
< 0\ ,\label{elliptic}
\end{equation}
for hyperbolic, parabolic and elliptic solutions respectively%
\footnote{However, near the origin, it is the sign of $RE/M$ rather
than $E$ that determines the type of solution.}.
Specification of the three arbitrary functions --- $M(r)$, $E(r)$
and $t_B(r)$ --- fully determines the model. They constitute a
radial coordinate choice, and two physical relationships.

\subsection{The past null cone (PNC)}

The notation and null cone solution used here were first developed
in \cite{MBHE}. However, they chose to work with the parabolic LTB
model, and hence, their gauge choice which locates the null cone of
the observer at one instant of time is simpler. This gauge choice was
generalised to all spatial sections, i.e. for all values of
$E$, in \cite{MHE}. In this latter paper, which we will call MHE,
they gave a complete outline of the observer's null cone in the LTB
model, and how one can relate the LTB model to observables using
this more general gauge choice. Therefore, we follow the general
outline given in MHE here.

On the one hand specification of the three arbitrary functions is
what determines the LTB model, and on the other the angular diameter
distance and the redshift space number density are what is given on
the observer's past null cone. Hence, we first need to locate the
null cone, and then relate the LTB arbitrary functions to the given
data.

Human observations of the sky are essentially a single event on
cosmological time scales, and as a result, being able to locate a single
null cone is all we need here; no general solution is needed. On
radial null geodesics, we have $ds^2 = 0 = d \theta^2 = d \phi^2$.
From (\ref{LTBmetric}), if the past null cone of the observation
event ($t = t_0, r = 0$) (here and now) is given by $t =
\hat{t}(r)$, then $\hat{t}(r)$ satisfies
\begin{equation}
d \hat{t} = - \frac{\ R'[\hat{t}(r),r]\ }{\sqrt{1 + 2E}}\ dr = -
\frac{\widehat{R^{'}}}{\ \sqrt{1 + 2E}\ }\ dr\ . \label{LTBmetricNC}
\end{equation}
We will denote a quantity evaluated on the observer's null cone, $t
= \hat{t}(r)$, by a $~\hat{}$\ ; for example $R[\hat{t}(r),r] \equiv
\widehat{R}$, and we note that it is a function of $r$ only instead
of $r$ and $t$. If we choose coordinate $r$ in such a way that, on
the past null cone of ($t_0,r$), we have
\begin{equation}
\frac{\widehat{R^{'}}}{\ \sqrt{1 + 2E}\ } = 1\ ,
\label{CoordinateChoice}
\end{equation}
then the incoming radial null geodesics are given by
\begin{equation}
\hat{t}(r) = t_0 - r\ . \label{MHE8}
\end{equation}
With our coordinate choice (\ref{CoordinateChoice}), the density
(\ref{density}) and the Friedmann equation (\ref{Rdot}) on the past
null cone then become
\begin{equation}
4\pi \hat{\rho} \hat{R}^2 = \frac{M^{'}}{\ \sqrt{1 + 2E}\ }\ ,
\label{MHE9}
\end{equation}
\begin{equation}
\widehat{\dot R} = \pm \sqrt{\frac{\ 2M\ }{\hat{R}} + 2E(r)}\ .
\label{MHE10}
\end{equation}
The gauge equation is then found from the total derivative of $R$ on
the null cone,
\begin{equation}
\frac{\ d \hat{R}\ }{dr} = \widehat{R^{'}} + \widehat{\dot{R}}
\frac{\ d \hat{t}\ }{dr}\ , \label{MHE11}
\end{equation}
and this, together with (\ref{CoordinateChoice}), (\ref{MHE8}) and
(\ref{MHE10})%
\footnote{Although we do not strictly know the sign of
$\widehat{\dot R}$, it is fairly safe to assume that it is positive
on our past null cone on the large scales that we are considering.
From now on we take the positive sign for the right hand side of
equation (\ref{MHE10}).}, leads to
\begin{equation}
\frac{\ d \hat{R}\ }{dr} - \sqrt{1 + 2E} = - \widehat{\dot{R}} = -
\sqrt{\frac{\ 2M\ }{\hat{R}} + 2E(r)}\ . \label{MHE12}
\end{equation}
We can then obtain an expression for $\sqrt{1 + 2E(r)}$,
\begin{equation}
W(r) \equiv \sqrt{1 + 2E} = \left \{\frac{1}{2} \left [ \left
(\frac{\ d \hat{R}\ }{dr} \right )^2 + 1 \right ] - \frac{\ M\
}{\hat{R}} \right \} \left / \left (\frac{\ d \hat{R}\ }{dr} \right
) \right.\ ,\label{MHE13}
\end{equation}
where a new variable $W(r)$ is introduced.
This expression tells us for which regions the spatial sections are
hyperbolic $1 + 2E > 1$, parabolic $1 + 2E = 1$ or elliptic $1 + 2E
< 1$, based on data obtained from the null cone. We substitute
(\ref{MHE13}) into (\ref{MHE9}) and rearrange it into the form
\begin{equation}
\frac{\ dM\ }{dr} + \left ( \frac{\ 4\pi \hat{\rho} \hat{R}\
}{\frac{\ d\hat{R}\ }{dr}} \right )M = \left ( \frac{\ 2\pi
\hat{\rho}\hat{R}^{2}\ }{\frac{\ d\hat{R}\ }{dr}} \right ) \left [
\left ( \frac{\ d\hat{R}\ }{dr} \right )^{2} + 1 \right ]\ .
\label{MHE14}
\end{equation}
The proper time from the bang surface to the past null cone
along the particle worldlines is described by
\begin{equation}
\tau(r) \equiv \hat{t}(r) - t_{B}(r) = t_{0} - r - t_{B}\ .
\label{MHE17}
\end{equation}

\subsection{Redshift formula}

Since the cosmological observations are given in terms of redshift
rather than the unobservable coordinate $r$, we need to express all
the relevant quantities in terms of redshift $z$. In order to do
this, the redshift formula is developed here.

As shown in MHE and elsewhere, the redhsift in LTB models is
\begin{equation}
\ln (1 + z) = \int^{r_{em}}_{0} \frac{\dot{R}^{'}(t,r)}{\ \sqrt{1 +
2E\ }\ }\ dr \label{MHE24}
\end{equation}
for the central observer at $r = 0$, receiving signals from an
emitter at $r = r_{em}$.

We need to find the redshift $z$ explicitly in terms of $r$,
$\hat{R}$ and $\hat{\rho}$, which we will later relate to
observables. We differentiate (\ref{Rdot}) with respect to $r$, then
evaluate it on the observer's past null cone, and we find
\begin{equation}
\frac{\widehat{\dot{R}^{'}}}{\ \sqrt{1 + 2E}\ } = \frac{1}{\
\widehat{\dot{R}}\ } \left[ \frac{M'}{\ \hat{R} \sqrt{1 + 2E}\ } -
\frac{M}{\ \hat{R}^2\ } + W' \right ]\ .\label{MHE26}
\end{equation}
From (\ref{MHE13}), we can get the derivative of $W$. Using equation
(\ref{MHE9}) to eliminate $M'$, and combining with equations
(\ref{MHE12}) and (\ref{MHE24}), it now follows that
\begin{equation}
\frac{\ dz/dr\ }{1 + z} = - \left ( \frac{\ d^2\hat{R}\ }{dr^2} + 4
\pi \hat{\rho} \hat{R} \right ) \left / \frac{\ d \hat{R}\ }{dr}
\right.\label{MHE29}
\end{equation}
with $z(r=0) = 0$. So theoretically we now have the redshift in terms
of coordinate radius $r$ from $\hat{R}(r)$ and $\hat{\rho}(r)$
directly if we integrate (\ref{MHE29}) with respect to $r$.

\subsection{The observables and the LTB model}

As explained, we are assuming a spherical metric with a central
observer purely for purely pragmatic reasons --- one does not tackle
the full generality of a complicated problem all at once. For
simplicity, we suppose there is only one type of cosmic source and
we only consider bolometric luminosities as in MHE. See Hellaby
\cite{Hellaby01} for a discussion of multiple source types and
multicolour observations. It is assumed that the luminosity and the
number density of each source can evolve with time; with the former
written as an absolute bolometric luminosity $L$, and the latter as
a mass per source, $\mu$. Isotropy about the Earth is assumed, and
we also assume that the universe is described by zero-pressure
matter - `dust', and galaxies or perhaps clusters of galaxies are
taken as the particles of this dust.

The two source evolution functions might naturally be expressed as
functions of local proper time since the big bang, $L(\tau)$ and
$\mu(\tau)$. However, one cannot be sure of the age of the objects at
redshift $z$ because the bang time is uncertain in a LTB model and
also because the location of the null cone is uncertain. The proper
time from bang to null cone will be a function of redshift,
$\tau(z)$, and the projections of the evolution functions on the
null cone are written as $\hat{L}$ and $\hat{\mu}$. Of course,
$\tau(z)$ is unknown until we have solved for the LTB model that
fits the data. For the sake of simplicity, we will take $\hat{L}$
and $\hat{\mu}$ to be given as function of $z$, and we use $l$ for the
apparent luminosity and $n$ for the number count observations. In
practice, many observational studies of evolution express their
results in terms of $z$.

The area distance $d_D$ (or equivalently the diameter distance
$d_A$) is the true linear extent of the source over the measured
angular size, which is by definition the same as the areal radius of
the source at the time of emission, i.e. $R$ in the LTB model. It
multiplies the angular displacements to give proper distance
tangentially and its projection onto the observer's null cone gives
the quantity $\hat{R}$. The luminosity distance is measurable if we
know the true absolute luminosity of the source at the time of
emission $\hat{L}$. If the observed apparent luminosity is $l(z)$,
then from the reciprocity theorem \cite{Ellis71} gives the
relationship between the diameter distance $d_A = \hat{R}$ and the
luninosity distance $d_L$\footnote{Note that equation (31) in MHE is
incorrect.},
\begin{equation}
(1 + z)^2 \hat{R}(z) = d_L(z) = \sqrt{\frac{\ \hat{L}\ }{l}}\
d_{10}\ , \label{MHE31}
\end{equation}
where $d_{10} = 10$~parsecs.

Let the observed number density of sources in redshift space be
$n(z)$ per steradian per unit redshift interval, hence the number
observed in a given redshift interval over the whole sky is
\begin{equation}
4 \pi n\ \delta z \ .\label{MHE33}
\end{equation}
Thus the total rest mass between $z$ and $z + \delta z$ is
\begin{equation}
4 \pi \hat{\mu}n\ \delta z\ ,\label{MHE34}
\end{equation}
where $\hat{\mu}(z) = \mu[\tau (z)]$ is the mean mass per source. Given the
local proper density on the null cone $\hat{\rho}$, the total rest
mass between $r$ and $r + dr$ evaluated on the null cone is
\begin{equation}
\hat{\rho} \widehat{d^3 V} = \hat{\rho}\ \frac{4 \pi \hat{R}^2
\widehat{R^{'}}}{\ \sqrt{1 + 2E}\ }\ dr\ , \label{MHE35}
\end{equation}
where $\widehat{d^3 V}$ is the proper volume on a constant time
slice. Hence, from (\ref{MHE34}),(\ref{MHE35}) and
(\ref{CoordinateChoice}), we get
\begin{equation}
\hat{R}^{2}\hat{\rho} = \hat{\mu}n\frac{dz}{dr}\ . \label{MHE36}
\end{equation}

\subsection{The differential equations}

Most of the equations developed above are given as differential
equations (DEs), and numerically DEs are easy to work with.  Therefore, we need a set of DEs, that
will generate the values of $r$, $M$, $E$ and hence $t_B$ from the
observations. The LTB model implied by the observations is thus deduced.

Transforming (\ref{MHE29}) to be in terms of redshift $z$ instead of
coordinate $r$, we obtain the null Raychaudhury equation
\begin{equation}
\hspace{-15 mm} \frac{\ d \hat{R}\ }{dz}\ \frac{\ d^2 z\ }{dr^2}\ (1
+ z) + \left [ \frac{\ d^2\hat{R}\ }{dz^2}\ (1 + z) + \frac{\ d
\hat{R}\ }{dz} \right ] \left( \frac{\ dz\ }{dr} \right )^2 = - 4\pi
\hat{\rho} \hat{R}(1 + z)\ . \label{MHE37}
\end{equation}
Substituting (\ref{MHE36}) into (\ref{MHE37}), using the facts
that $\frac{dr}{\ dz\ } = 1 \left / \frac{dz}{\ dr\ } \right.$ and $
\frac{\ d^2 r\ }{dz^2} = - \left ( \frac{\ dr\ }{dz} \right )^3
\frac{\ d^2z\ }{dr^2}$, rewriting it so that all terms involving
$\hat{R}$ (and hence $d\hat{R}/dz$ and $d^2\hat{R}/dz^2$) are on one
side of the equation, we then have a second order DE for $r(z)$:
\begin{eqnarray}
\frac{\ d^2r\ }{dz^2} = \left [ \frac{\ d\hat{R}\ }{dz}\ (1 + z)
\right ]^{-1} \left \{ \left [ \frac{\ d^2\hat{R}\ }{dz^2}\ (1 + z)
+ \frac{\ d\hat{R}\ }{dz} \right ] \left ( \frac{\ dz\ }{dr} \right
) \right. \nonumber \\ \hspace{10 mm} \left. + \frac{\ 4
\pi\hat{\mu}n\ }{\hat{R}}\ (1 + z) \right \}\left( \frac{\ dr\ }{dz}
\right)^2\ .\label{secondrofz}
\end{eqnarray}
Since we want to solve all our DEs (and hence get the values for our
functions $r$, $M$ and $E$) in parallel, we need to introduce a new
variable such that we can rewrite (\ref{secondrofz}) as two first
order DEs.

We introduce a new variable $\phi = \phi(z)$, defined by
\begin{equation}
\frac{\ dr\ }{dz} = \phi\ , \label{DE1}
\end{equation}
and equation (\ref{secondrofz}) then becomes
\begin{equation}
\frac{\ d \phi\ }{dz} = \left \{ \frac{1}{\ 1 + z\ } + \frac{\
\frac{\ d^2 \hat{R}\ }{dz^2} + \frac{\ 4 \pi \hat{\mu}n \phi\
}{\hat{R}}\ }{\frac{\ d \hat{R}\ }{dz}} \right \} \phi\ .
\label{DE2}
\end{equation}
If we transform (\ref{MHE13}) into a function of $z$ and take square
root of both sides, using the inverse of (\ref{DE1}) we then obtain
\begin{equation}
W = \sqrt{1 + 2E\ } = \frac{\ \frac{\ d\hat{R}\ }{dz}\ }{2\phi} +
\frac{\ \left (1 - \frac{\ 2M\ }{\hat{R}} \right )\phi\ }{2\frac{\
d\hat{R}\ }{dz}}\ . \label{W}
\end{equation}
We substitute (\ref{MHE36}) into (\ref{MHE14}) and rewrite it as
$dM/dz$ instead of $dM/dr$, so together with equation (\ref{W}) we
then get
\begin{equation}
\frac{\ dM\ }{dz} = 4 \pi \hat{\mu}n \sqrt{1 + 2E\ }\ . \label{DE3}
\end{equation}

Hence, equations (\ref{DE1})-(\ref{DE3}) give us a set of coupled
first order DEs that we use in order to generate the values for
$r(z)$, $M(z)$ and $W(z)$ (or $E(z)$) from the observational data.
We can then obtain the values for $\hat{\eta}$, $\tau$ and then the
third arbitrary function $t_B(z)$ for the hyperbolic and elliptic
cases by substituting these values into equations
(\ref{hyperbolic}), (\ref{elliptic}) and (\ref{MHE17}), with
(\ref{hyperbolic}) and (\ref{elliptic}) evaluated on the null cone.
However, there is a borderline case -- the near-parabolic case, which is
needed where the exact expressions become numerically intractable.  See
\ref{NearPara} for the details. Note that from
equation (\ref{DE2}), if we know the values for $\hat{R}$ and $4 \pi
\hat{\mu}n$, we can then solve for $\phi$ independently without
knowing the values of $r$, $M$ and $W$; while solving for $r$, $M$
and $W$ depends on knowing $\phi$. This property of the $\phi$ equation
will be very useful later on.

\subsection{Origin conditions}

At the origin of spherical coordinates, $r = 0$, we have $R(t,0) =
0$ and $\dot{R}(t,0) = 0$ for all $t$. Hence, on the observer's past
null cone equations (\ref{MHE11}) and (\ref{MHE12}) then become
\begin{equation}
\left. \frac{d \hat{R}}{dr} \right |_{r = 0}  = \widehat{R^{'}} =
\sqrt{1 + 2E} = 1\ ,\label{MHE20}
\end{equation}
and thus $\hat{R} \approx r$ to lowest order near $r = 0$. From
(\ref{MHE9}) we then find that
\begin{equation}
M' \approx 4 \pi \hat{\rho}_0 r^2\ ,\ \ \ M \approx \frac{\ 4\ }{3}\
\pi \hat{\rho}_0 r^3\ ; \label{MHE21}
\end{equation}
and from (\ref{MHE13}) using a Taylor series for $\hat{R}$, and
working to second order in $r$, we get
\begin{equation}
E \approx  \left [ \frac{1}{2}\left (\frac{d^2 \hat{R}}{dr^2}
\right)_0^{2} - \frac{\ 4\ }{3}\ \pi \hat{\rho}_0 \right ] r^2\ ,
\label{MHE22}
\end{equation}
where $d^2 \hat{R}/dr^2$ is finite when $r = 0$. Note that equations
(\ref{MHE21}) and (\ref{MHE22})\footnote{Note that equation (22) given in MHE is
incorrect. Their expression does not allow all three cases of $E$.}
give us $M \propto E^{3/2}$.

We can get the origin limit for $\hat{\mu}n$ from equation
(\ref{MHE36}), which is
\begin{equation}
\hat{\mu}n \approx \frac{r^2 \hat{\rho}_0}{\ \left. \frac{d^2
\hat{R}}{dz^2} \right|_0\ }\ .
\end{equation}
Also, if we substitute (\ref{MHE20}) into (\ref{MHE29}), after
rearranging the expression, we then get the origin condition for
$dz/dr$ (and hence $z$)
\begin{equation}
\frac{dz}{\ dr\ } \approx \left. \frac{\ d^2 \hat{R}\ }{dz^2}
\right|_0 \ .
\end{equation}

The behaviour of the DEs (\ref{DE1})-(\ref{DE3}) needs to be checked
near the origin, i.e. $z \rightarrow 0$. Since $r$ and $z$ have a
linear relation, we know that near the origin, $\hat{R} \sim z$, $4
\pi \hat{\mu}n \sim z^2$, $d\hat{R}/dz$ is finite, $d^2
\hat{R}/dz^2 = \mbox{finite}$ and $M \sim z^3$. Also we know that
$dz/dr(0) = 1/(d\hat{R}/dz(0))$, so $\phi = \mbox{finite}$. Hence,
our DEs are well behaved as $z \rightarrow 0$.

\subsection{Apparent horizon and the maximum in $\hat{R}$}

In the early universe, the expansion is so rapid that the light rays
that are headed towards us are actually getting further away. One
can consider the set of photons that are all the same time away from
observation to be an incoming wavefront. As the universe slows down,
there comes a moment when the area of such a wavefront is
stationary, and $R(t,r)$ has reached its maximum value. The locus of
such points for all incoming wavefronts is the apparent horizon.
Hence, for the LTB model, the maximum of the areal radius (or
diameter distance) down the null cone is where the null cone crosses
the apparent horizon. We locate this point by the calculation below.

Since the apparent horizon is the hypersurface in spacetime where
$\hat{R}$ is momentarily constant, we put $d \hat{R}/dr = 0$ into
(\ref{MHE11}) and using (\ref{CoordinateChoice}), (\ref{MHE8}) and
(\ref{MHE10}), we get
\begin{equation}
\sqrt{\frac{\ 2M\ }{\hat{R}} + 2E} = \sqrt{1 + 2E}\ ,
\end{equation}
and hence
\begin{equation}
\hat{R} = 2M\ . \label{Rover2M}
\end{equation}
We will see that this locus presents us with a particular difficulty
in our numerical reduction of null cone data. Of course, in the case
when the cosmological constant is not set to zero, and if we are
considering both the future and the past horizon; the calculation
and the analysis will be more complicated \cite{H87,KraHe04}. See
also \cite{Hellaby06} for the observational significance of this
locus.

There are a few things worth considering here -- our DEs become
singular when we reach the maximum in the areal radius (diameter
distance) $\hat{R}$, i.e. $d\hat{R}/dz = 0$. From equation
(\ref{Rover2M}) we know that at the maximum of $\hat{R}$, we have
$\hat{R} = 2M$. If one looks at equations (\ref{W}) and (\ref{DE3}),
it actually contains zero over zero at this point, and any numerical
method will break down here. Further, from (\ref{MHE29}) and
(\ref{DE2}) we can see that $d^2 \hat{R}/dz^2 = -4 \pi
\hat{\mu}n/\hat{R}$ where $d\hat{R}/dz = 0$. There is no problematic
behaviour of $\phi$ here, as can be verified in the FLRW case.
Hence, in order to carry our numerics through the maximum of
$\hat{R}$, we need to perform a series expansion near this point for
$\hat{R}(z)$, $\hat{\mu}n(z)$, $\phi(z)$, $M(z)$ and $W(z)$, as given
in (\ref{Seriesmn})-(\ref{SeriesW}) of \ref{RmaxSeries}.

Here, we use $\hat{R}_{max}$ to denote the maximum in $\hat{R}$, and
its corresponding $z$ value is called $z_{m}$. The series are then
written in powers of $\Delta z = z - z_m$. From the $\hat{R}$ and
$\hat{\mu}n$ data, we can easily determine the values of
$\hat{R}_{max}$, $\hat{\mu}n(z_m)$ and $z_{m}$, and thus the
remaining $\hat{R}(z)$ and $\hat{\mu}n(z)$ coefficients can be
evaluated by simply performing a least squares fit using the data
values near $z_{m}$. In order to obtain the expressions for the
coefficients in the $\phi(z)$, $M(z)$ and $W(z)$ series, we need to
substitute (\ref{Seriesmn})-(\ref{SeriesW}) from \ref{RmaxSeries}
into our DEs (\ref{DE1})-(\ref{W}). The detailed expressions can be
found in \ref{RmaxSeries}.

From (\ref{Phi0})-(\ref{Phi3}) we can see that all $\phi(z)$
coefficients are determinable once we know the values of $z_{m}$ and
all coefficients of the $\hat{R}$ and $4 \pi \hat{\mu}n$ fits. Using
(\ref{DE1}) and (\ref{SeriesPhi}), the series expansion for $r$ is
simply
\begin{equation}
r(z) = r_{0} + \phi_{0}\Delta z + \frac{\ \phi_{1}\ }{2}\ \Delta z^{2} + \frac{\
\phi_{2}\ }{3}\ \Delta z^{3} + \frac{\ \phi_{3}\ }{4}\ \Delta z^{4} + \cdots\ ,
\label{Seriesr}
\end{equation}
where $r_{0} = r(z_m)$ is the integration constant.

Note that $M_0$ is obtained directly from $\hat{R}_{max}$ without
any further information. The only problem is that
expressions (\ref{M2}), (\ref{M3}) and (\ref{W0})-(\ref{W2}) in
\ref{RmaxSeries}, all depend linearly on $M_1$. Unfortunately, no
information about $M_1$ can be obtained when we carry out the series
expansion, as one can see from (\ref{M1}). Despite this, it is still
possible to obtain a value for $M_1$ by substituting a known value (from
numerical integration), say $M_{a}$ at $z_{a}$, where $z_{a}$ is some distance away from
$z_{m}$, into (\ref{SeriesM}) as described in more detail below.

\section{Numerical procedure}

\subsection{Data handling and numerical method}

The actual data we must use consists of redshift and apparent magnitude
measurements for a large number of discrete sources, which must be
sorted into redshift bins.  In each bin we must calculate the total
number of sources, $4 \pi n \delta z$, and the average value for the diameter
distance, $\hat{R}$.

Now the above theory treats all physical and geometric quantities,
such as the density and the metric, as continuous functions of
position, while the available data is a discrete set of sources.
Therefore it might at first seem one should fit a smooth curve to
the data in order to proceed with the integration.  However,
numerical methods are not continuous either, and any numerical
method we might choose to solve our PNC equations with is based on
discretisation of continuous DEs. Furthermore, we must be careful
not to hide any inhomogeneity by smoothing on too large a scale, or
introduce unintended bias by inappropriate choice of smoothing
function.  We note that the process of calculating averages on the
redshift bins already introduces a measure of smoothing and a basic
smoothing scale.  We also note that higher order integration methods
that use data from several different $z$ values will also have a
smoothing effect, and this is an option we are keeping open.  So
although statistical fluctuations in the real data may need a
further degree of smoothing, we prefer to keep it to a minimum, only
introducing as much as necessary.  We also argue that any extra
smoothing that is required should be closely tied to the numerical
integration scheme, and not merely ad hoc.

An important consideration in the choice of numerical method is that
the right hand sides of the DEs (\ref{DE1})-(\ref{DE3}) contain not
only the funtions being solved for, $\phi(z)$, $r(z)$, $M(z)$ and
$E(z)$, but also the given data derived from observations,
$\hat{R}(z)$ and $4 \pi \mu n(z)$. Since the latter are only known
at discrete $z$ values (the mid points of the $z$ bins), methods
that allow adaptive step sizes are not appropriate, and similarly
evaluations of $\phi(z)$, $r(z)$, $M(z)$ and $E(z)$ significantly
above or below their correct values should be avoided because there
is no way to find the corresponding values of the given data.  A
second consideration is that with real observations, there will be
statistical fluctuations and measurement uncertainties, so there is
a limit to how much improvement can be gained from using higher
order methods.  In line with our policy of not using a more
complicated method than the situation demands, we found that, with
bin size (= step size) $\delta z = 0.001$, a second order
Runge-Kutta method gave entirely satisfactory results when very
accurate fake data was given for the data functions.  These choices
may change in the future once real data is used and as more factors
are included.

Once $\phi(z)$, $r(z)$, $M(z)$ and $E(z)$ are determined, then
$\hat{\eta}(z)$, $\tau(z)$ and $t_B(z)$ are easily obtained from the
algebraic equations (\ref{MHE8}) and
(\ref{hyperbolic})-(\ref{elliptic}). However, at each discrete
position, we are required to determine, numerically, which type of
evolution we have: hyperbolic, elliptic or near-parabolic. Note that
for the near-parabolic case, when $E$ is small but not zero, we use
equation (\ref{tauhyperbolicnearparabolic}) from \ref{NearPara}. As
one might have noticed, equation (\ref{tauhyperbolicnearparabolic})
is in powers of $\hat{R} (2E)/M$, and this factor can be evaluated
at each discrete position since we know the values for $\hat{R}$,
$E$ and $M$. Of course, the error for this approximation of the
series expansion for $\tau$ has to be small, say about $10^{-7}$; if
we take (\ref{tauhyperbolicnearparabolic}) up to order 3, this will
give us $\left|\frac{\ \hat{R}(2E)\ }{M}\right| \approx 0.1$. Hence,
if $\frac{\ \hat{R}(2E)\ }{M} > 0.1$, we use the hyperbolic case
(\ref{hyperbolic}); if $\frac{\ \hat{R}(2E)\ }{M} < -0.1$, we use
the elliptic case (\ref{elliptic}), and if $-0.1 < \frac{\
\hat{R}(2E)\ }{M} < 0.1$ we use the near-parabolic case.

A set of computer programmes\footnote{Matlab was used
for our numerical work.} were developed that
generate the values for the LTB functions $M$, $E$ and $t_{B}$.
Below we give a brief summary of the order of the steps followed in
our programmes. To obtain the mass, energy and bang time functions
($M$, $E$ and $t_{B}$ respectively) from observational data and
source evolution, we proceed as follows:

(i) take the discrete observed data for $l(z,\theta,\phi)$ and
$n(z,\theta,\phi)$, divide it into redshift bins of chosen width
$\delta z$, and in each bin average or sum it over all angles and the bin width
to obtain $l(z)$ and $n(z)$. We may wish first to correct the data
for known distortions and selection effects due to proper motions,
absorption, shot noise, image distortions, etc.;

(ii) choose evolution functions $\hat{L}(z)$ and $\hat{\mu}(z)$ based
on whatever observations and theoretical arguments may be mustered;

(iii) determine $\hat{R}(z)$ from $\hat{L}(z)$ and $l(z)$ using
(\ref{MHE31}), this is then our first input data function and we
have $4 \pi \hat{\mu}n$ as our second input data function;

(iv) numerically integrate the DEs (\ref{DE1})-(\ref{DE3}) using the
redshift bins as the basic step size, and the binned data for $z$,
$\hat{R}$ and $4 \pi \hat{\mu}n$, thus obtaining $r(z)$, $M(z)$ and
$E(z)$;

(v) solve for $\hat{\eta}$, $\tau(r)$, and hence $t_{B}(r)$ from
(\ref{hyperbolic})-(\ref{elliptic}) and (\ref{MHE17}), with
(\ref{hyperbolic})-(\ref{elliptic}) evaluated on the null cone.
Notice that $L(\tau)$ and $\mu (\tau)$ could also be found in this
step.

 However, at this early stage of development, steps (i)-(iii) use test data
generated from a variety of model assumptions.  In fact, step (iv) has four
components, which are summarised below:
\begin{itemize}
\item   Deduce the origin parameters and output for the first 3 data
points, i.e. at $z = 0$, $z = (1/2) \delta z$ and $z = (3/2) \delta z$.

\item   Use numerical DE solvers - a second order Runge-Kutta method
- for solving the DEs up to just before the maximum in $\hat{R}$ is reached.

\item   Determine the point $z_a$ where the switch to the series expansion
is made, evaluate all quantities to be matched such as $M_a$, $W_a$, etc,
and extend the numerics through the maximum in $\hat{R}$ by calculating
the series expansions of \ref{RmaxSeries} for $r$, $\phi$, $M$ and $E$.

\item   Evaluate another matching value for switching back from the
series expansion to numerical integration, and continue to solve the
DEs numerically up to the limit of the data, here set to $z = 3$.

\end{itemize}

\subsection{Data near the origin}

Although we have already discussed how the DEs behave near the origin,
from any available cosmological data that we might use, there is no
data available at the origin itself, and very little in the first
few redshift bins. Therefore, a method of filling in this gap is
necessary in order to provide the initial values needed by the
numerical integration.

If we average over all the data values within each bin, for example
the $\hat{R}$ values within a given $z$ bin, then the average
$\hat{R}$ values that we use are located roughly in the middle of each bin.
Thus we have the first value of $\hat{R}$ at $z = \delta z/2$. We
can then get the discretised versions of $d\hat{R}/dz$ and
$d^{2}\hat{R}/dz^{2}$ from the first and second differences of
$\hat{R}$. It takes two $\delta z$ bins to get the first value of
the first difference at $z = \delta z$ and all the values are
located at $z = k \delta z$ for any positive integer $k$. However,
it takes three $\delta z$ bins to get the first value of the second
difference and hence the first value of $d^{2}\hat{R}/dz^{2}$ at $z
= 3\delta z/2$, therefore, all the values are located in the middle
of each bin.

Since we want to have a complete set of data at each $z$ value, we
take the average of the two neighbouring first difference data
points to get all our data at the half $\delta z$ locations (in the
middle of each bin). In doing so, we will not have data values at
the origin or at $z = \delta z/2$, since the first complete data set
is at $z = 3/2 \delta z$. But we know that LTB is RW like near the
origin due to the fact that it assumes spherical symmetry. For that
reason, the series expansions of the RW expressions are used for
finding the RW parameters that fit the data values at the origin and
at $z = \delta z/2$, i.e. we determine the central values of $H_0$
(Hubble constant) and $q_0$ (deceleration parameter) from the data
near $z = 0$. So we take the standard RW expressions for
$\hat{R}(z)$ and $4 \pi \hat{\mu}n(z)$ given by equations
(A.1) and (A.2) in Appendix A of MHE, and we do series expansions
of them near the origin, as detailed in \ref{NearOrigin}.
The origin limits are $r(0) = 0$, $\phi(0) =
d\hat{R}/dz(0) = 1/H_0$, $M(0) = 0$, $E(0) = 0$, $\hat{R}(0) = 0$
and $d^2 \hat{R}/dz^2(0) = -(3 + q_0)/H_0$.

\subsection{Passing through the maximum in $\hat{R}$}

As mentioned before, a series approximation is required in the
vicinity of $\hat{R}_{max}$, where the DEs become singular.  Here
all $\phi(z)$ coefficients are determinable once we know the values
of $z_{m}$ and all coefficients of $\hat{R}$ and $4 \pi \hat{\mu}n$.
A set of $\phi$ values can be generated from the series expansion by
substituting a set of $\Delta z$ values into (\ref{SeriesPhi}), for
a $z$ interval that overlaps with our numerical results. We use the
$\hat{R}$ and $\hat{\mu}n$ values of 180 redshift bins on either
side of $z_{m}$ (this is a redshift interval of 0.361 which covers
about 12\% of the total redshift interval that we are considering
here), and perform a least squares fit with these data to obtain all
the coefficients for $\hat{R}$ and $\hat{\mu}n$, and hence obtain
all coefficients for $\phi$. There is good agreement between the
numerical and series values over a range of $z$ values when plotted
on the same graph, and there is one intersection point between the
two curves before $z_{m}$ (and also the closest to $z_{m}$). The
intersection points here are important since they are where we match
values between the series expansion and the numerical integration
for $\phi$.  This is what we choose for $z_a$, and the numerically
derived $M$ at $z_a$ becomes our $M_a$.

Now that we know where $z_a$ is, we can get $M_1$ from $z_a$ and
$M_a$ if we substitute them into equation (\ref{SeriesM}), using
$\Delta z = z_a - z_m$
\begin{equation}
M_{a} = M_{0} + M_{1}(z_{a} - z_{m}) + M_{2}(z_{a} - z_{m})^{2} +
M_{3}(z_{a} - z_{m})^{3} + \cdots\ . \label{SeriesMa}
\end{equation}
where $M_2$ and $M_3$ are given by (\ref{M2}) and (\ref{M3}).
Similarly, if we are matching the $W$ values
\begin{equation}
W_{a} = W_{0} + W_{1}(z_{a} - z_{m}) + W_{2}(z_{a} - z_{m})^{2} +
\cdots\ , \label{SeriesWa}
\end{equation}
where $W_1$ and $W_2$ are given by (\ref{W1}) and (\ref{W2}).
Therefore, if we match $M$ then
\begin{eqnarray}
  \fl M_1 & = \left \{ \frac{\ M_a - M_0\ }{z_a - z_m} +
\frac{\ 8 \pi^2 (\hat{\mu}n)_0^{\ 2}\ }{\hat{R}_{max}}\ (z_a - z_m) +
\frac{\ R_2\ }{2}\ (z_a - z_m)  \right. \nonumber \\
  \fl & + \left (\frac{R_2}{\ 1 + z_m\ } + \frac{\ (\hat{\mu}n)_1R_2\
}{(\hat{\mu}n)_0} + R_3 \right )\frac{\ (z_a - z_m)^2\ }{4}
\nonumber \\
  \fl & \left. + \frac{\ 8 \pi^2 (\hat{\mu}n)_0 (\hat{\mu}n)_1\
}{\hat{R}_{max}}\ (z_a - z_m)^2 \right \} \left / \left \{ 1 + \left
(\frac{1}{\ 1 + z_m\ } + \frac{\ (\hat{\mu}n)_1\ }{(\hat{\mu}n)_0}
\right )\frac{\ z_a - z_m\ }{2} \right. \right.
\nonumber \\
  \fl & \left. + \left (\frac{(\hat{\mu}n)_2}{\ 3(\hat{\mu}n)_0\ } +
\frac{(\hat{\mu}n)_1}{\ 3(\hat{\mu}n)_0(1 + z_m)\ } - \frac{R_2}{\
3\hat{R}_{max}\ } \right )(z_a - z_m)^2 \right \} \ .\label{M1fromM}
\end{eqnarray}
Alternatively, if $W$ is used for the matching, we have
\begin{eqnarray}
\fl M_1 & = \left \{ W_a + \frac{R_2}{\ 4 \pi (\hat{\mu}n)_0\ }\
(z_a - z_m) + \frac{\ 4 \pi (\hat{\mu}n)_0\ }{\hat{R}_{max}}\ (z_a -
z_m) \right.
\nonumber \\
\fl & + \frac{3R_3}{\ 16 \pi (\hat{\mu}n)_0\ }\ (z_a - z_m)^2 -
\frac{(\hat{\mu}n)_1R_2}{\ 16 \pi (\hat{\mu}n)_0^{\ 2}\ }\ (z_a -
z_m)^2 \nonumber \\
\fl & \left. + \frac{3R_2}{\ 16 \pi (\hat{\mu}n)_0(1 + z_m)\ }\ (z_a
- z_m)^2 + \frac{2 \pi (\hat{\mu}n)_1}{\hat{R}_{max}}\ (z_a - z_m)^2
\right\} \Bigg/
\nonumber \\
\fl & \left \{ \frac{1}{\ 4 \pi (\hat{\mu}n)_0\ } + \frac{z_a -
z_m}{\ 4 \pi (\hat{\mu}n)_0(1 + z_m)\ } - \frac{R_2(z_a - z_m)^2}{\
4 \pi (\hat{\mu}n)_0 \hat{R}_{max}\ } \right \}\ . \label{M1fromW}
\end{eqnarray}
All our functions should connect the numerical integration ($z <
z_a$) and series expansion ($z > z_a$) parts at $z_a$; and from
$z_a$, we can generate the corresponding $r_a$ and $M_a$ easily.
With all $\phi$ coefficients known, a value for $r_0$ can be found
from (\ref{Seriesr}). Using (\ref{M1fromM}) or (\ref{M1fromW}), a
value for $M_1$ can easily be determined.

The purpose of doing a series expansion is to extend our numerics
through $\hat{R}_{max}$, but once this is achieved, we need
to switch back to numerical integration again. It is sensible if we
connect at $z_{J}$ where $z_m - z_a = z_{J} - z_m$.

Initially, we tried matching $M$ at the two connecting points, $z_a$ and $z_J$.
However, after we compared $r$, $\phi$, $M$ and $W$ from our
numerics with the correct curves generated from the assumed model,
the $r$, $\phi$ and $M$
curves showed good agreement, but the $W$ curve had jumps at
the two connecting points. As anticipated, $W$ is the
least well-determined function.

We then tried to match $W$ at both connecting points.
Although this removed the two jumps in the numerical $W$ curve, it
also reduced the accuracy of the $W$ series expansion.
A key consideration is that at $z_m$ we actually know the value of $M$ from
(\ref{M0}) if we know $\hat{R}_{max}$. In order to maximise the
accuracy of our series expansion and minimise the jumps that appear
in our $W$ graph, we matched $M$ at the first connecting
point, and $W$ at the second one.

This approach does not leave any visible alteration in the $M$
curve, the jump in $W$ at the first connecting point is still present,
but better accuracy for the series expansion is obtained and the
second jump is avoided. One thing worth mentioning here is
that we may need to shorten the $z$ interval for the series
expansion, since with inhomogeneous data, fluctuations will be
present, so if the interval is too wide compared with the
fluctuations, the accuracy for our series expansion will be lower.
However, this problem will only be dealt with when it has shown a
significant effect on the numerics.

\section{Testing the numerics}

In order to test our numerical procedure, we need fake
``observational data" for which the correct results are known.
Therefore we generated sets of ``observational data" that would be
produced in a selection of LTB universes.

Although we did a full comparison of numerical output from our
programme, $M$, $E$ and $t_B$, with the correct LTB functions for a
variety of different models, both homogeneous and inhomogeneous, we
cannot present all our results here. Therefore, we summarise the
ones we did in Table \ref{summary_for_all_cases} below and only
present a complete set of plots from one homogeneous and one
inhomogeneous model in the two subsections below. In order to avoid
confusion, we call the $H_0$ and $q_0$ used for generating fake data
$q_{0d}$ and $H_{0d}$; and the ones our numerical procedure extracts
from the data $q_0$ and $H_0$ from here on. Where the model is
inhomogeneous, both pairs are the values at the origin, as explained
in Section 3.2.

\begin{table}[h]
\caption{Summary of all the homogeneous and inhomogeneous cases used
for full comparisons between our numerics and the generated fake
data.} \label{summary_for_all_cases}
\begin{center}
\hspace{.6in}
\begin{tabular}{|c|c|c|} \hline %
\multicolumn{3}{|c|}{Homogeneous cases}  \\ \hline
Hyperbolic & Near  & Elliptic  \\
($k = + 1$) & parabolic & ($k = - 1$)   \\ \hline
$q_{0d} = 0.45$ & $q_{0d} = 0.49$ & $q_{0d} = 0.8$   \\
$H_{0d} = 0.72$ & $H_{0d} = 0.72$ & $H_{0d} = 0.72$ \\ \hline
$q_{0d} = 0.1$ & $q_{0d} = 0.51$ & \\
$H_{0d} = 0.72$ & $H_{0d} = 0.72$ & \\ \hline
\multicolumn{3}{|c|}{Inhomogeneous cases} \\ \hline
Hyperbolic & Near & Elliptic  \\
& parabolic &  \\ \hline
\mbox{Varying bang time} & \mbox{Varying geometry or} &
\mbox{Strongly inhomogeneous} \\
\mbox{with}\ $q_{0d} = 0.2$ & \mbox{energy with}\ $q_{0d} = 0.52$ &
\mbox{with}\ $q_{0d} = 0.6$ \\ \hline
\mbox{Varying mass} & & \\
\mbox{with}\ $q_{0d} = 0.22$ & & \\ \hline
\end{tabular}
\end{center}
\end{table}

\subsection{Homogeneous models}

Amongst the homogeneous models we found that the near-parabolic models to have
slightly lower accuracy.  As noted above, we expect the output function with
largest error to be $W = \sqrt{1 + 2 E}$.

\begin{figure}[!h]
\includegraphics[width=10.4cm, height=7.65cm]{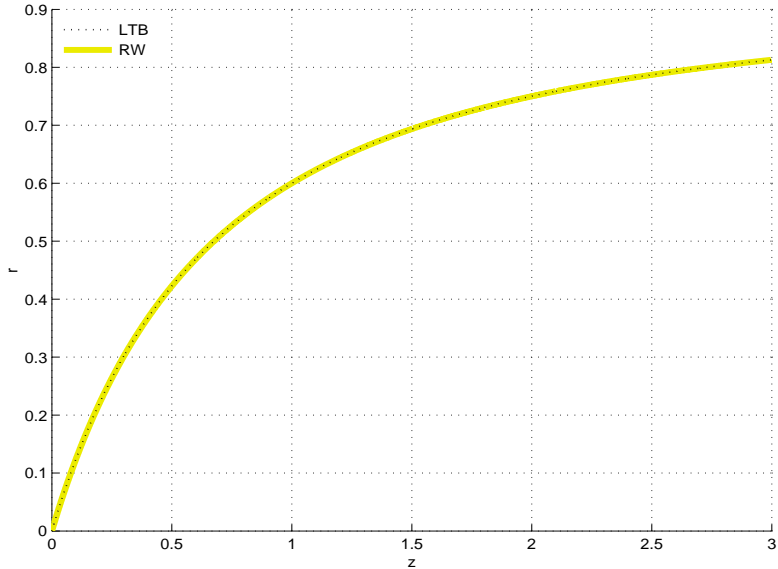}
\caption{Results of $r$ vs. $z$ with $H_{0} \approx 0.71999$, $q_{0}
\approx 0.490004$ and $\delta z = 0.001$. The grey curve is the
correct RW expression and the dotted black one is our numerical
output using Runge-Kutta as the integration method.} \centering
\label{home1_r}
\end{figure}

\begin{figure}[!t]
\includegraphics[width=10.4cm, height=7.65cm]{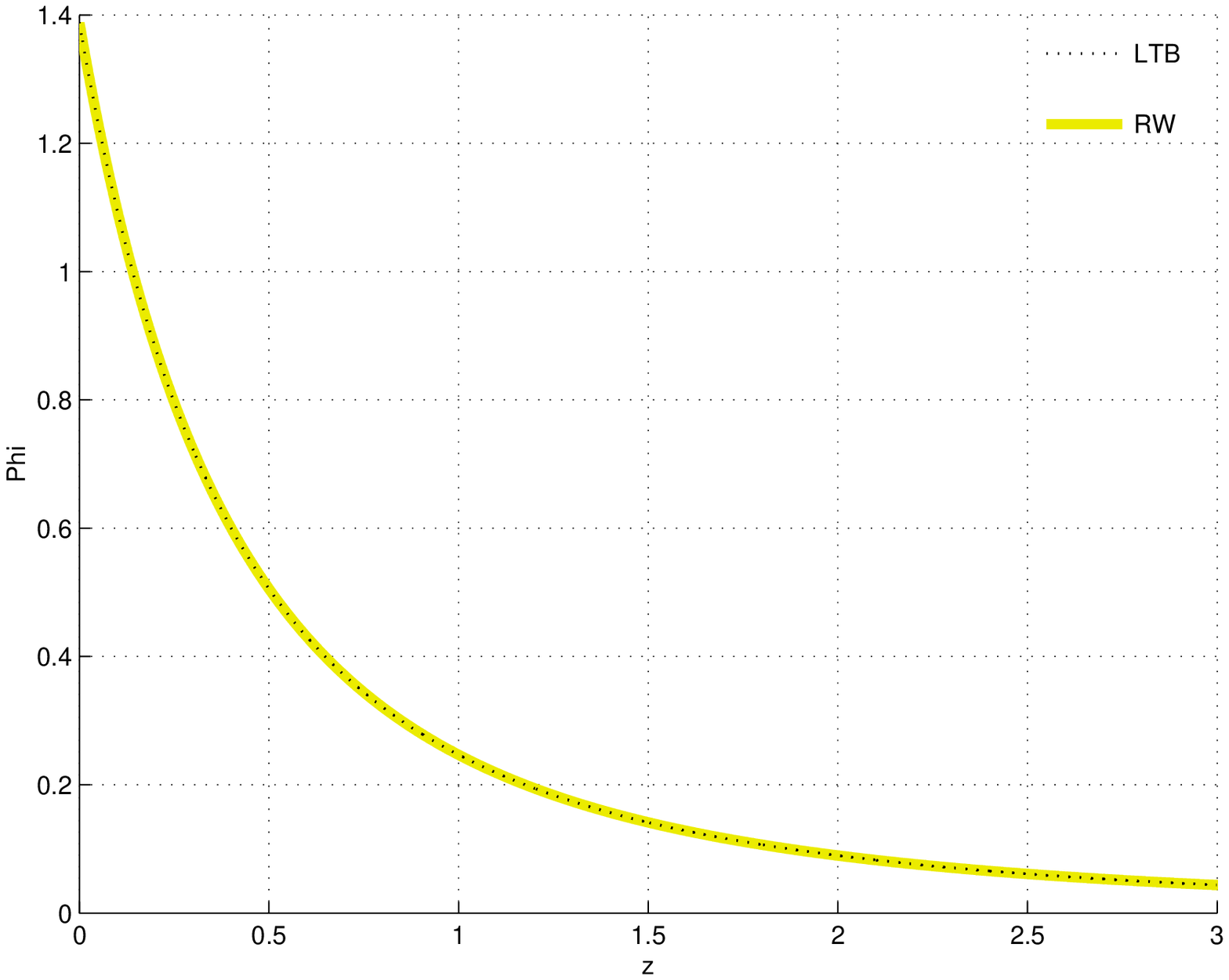}
\caption{Results of $\phi$ vs. $z$ with $H_{0} \approx 0.71999$,
$q_{0} \approx 0.490004$ and $\delta z = 0.001$. The grey curve is
the correct RW expression and the dotted black one is our numerical
output using Runge-Kutta as the integration method.} \centering
\label{home1_phi}
\end{figure}

\begin{figure}[!b]
\includegraphics[width=10.4cm, height=7.65cm]{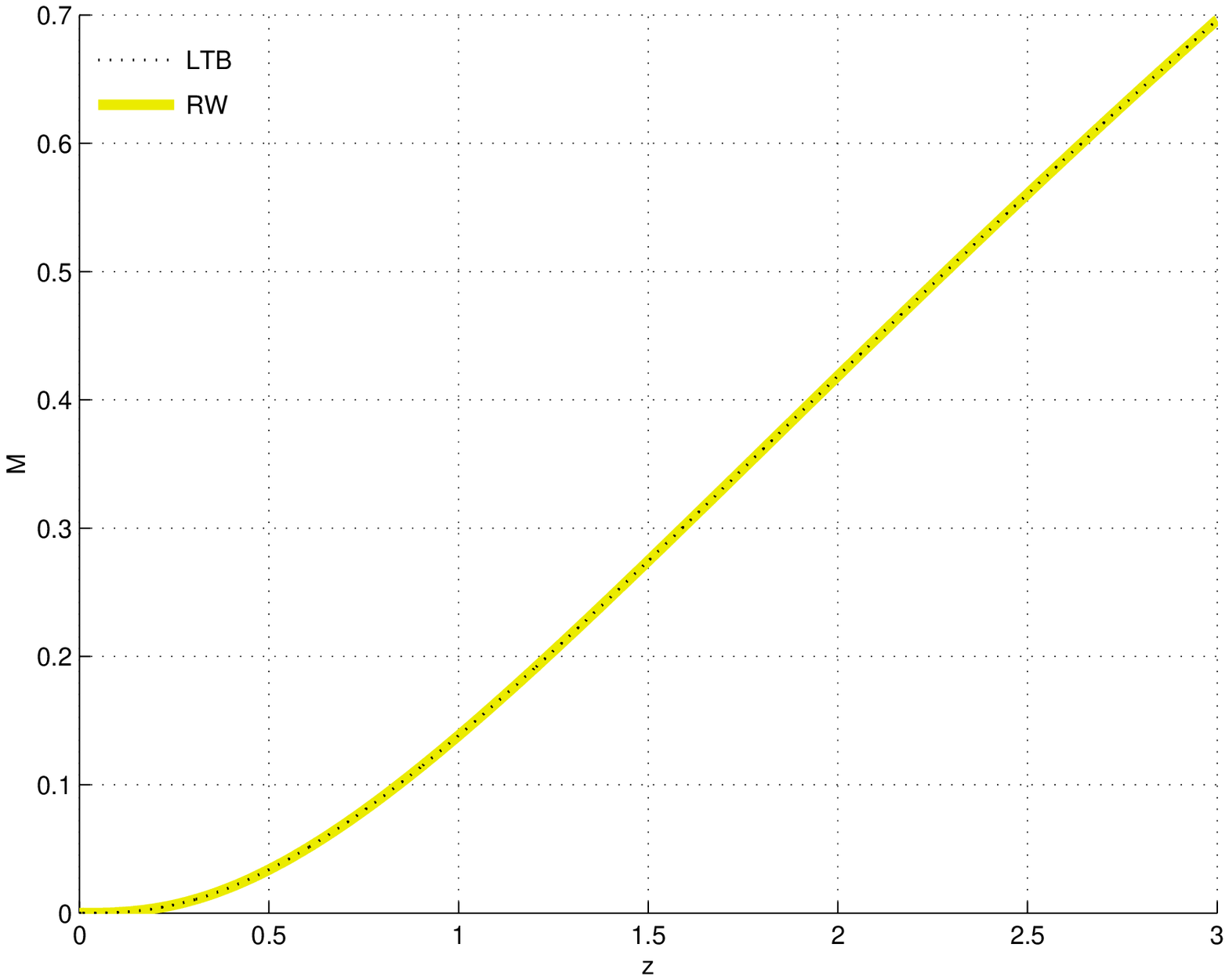}
\caption{Results of $M$ vs. $z$ with $H_{0} \approx 0.71999$, $q_{0}
\approx 0.490004$ and $\delta z = 0.001$. Matching the $M$ values at
the first connection point and the $W$ values at the second
connection point. The solid grey curve is the correct RW expression
and the dotted black one is our numerical output using Runge-Kutta
as the integration method.}
\centering \label{home1_M}
\end{figure}

\begin{figure}[!t]
\includegraphics[width=10.4cm, height=7.65cm]{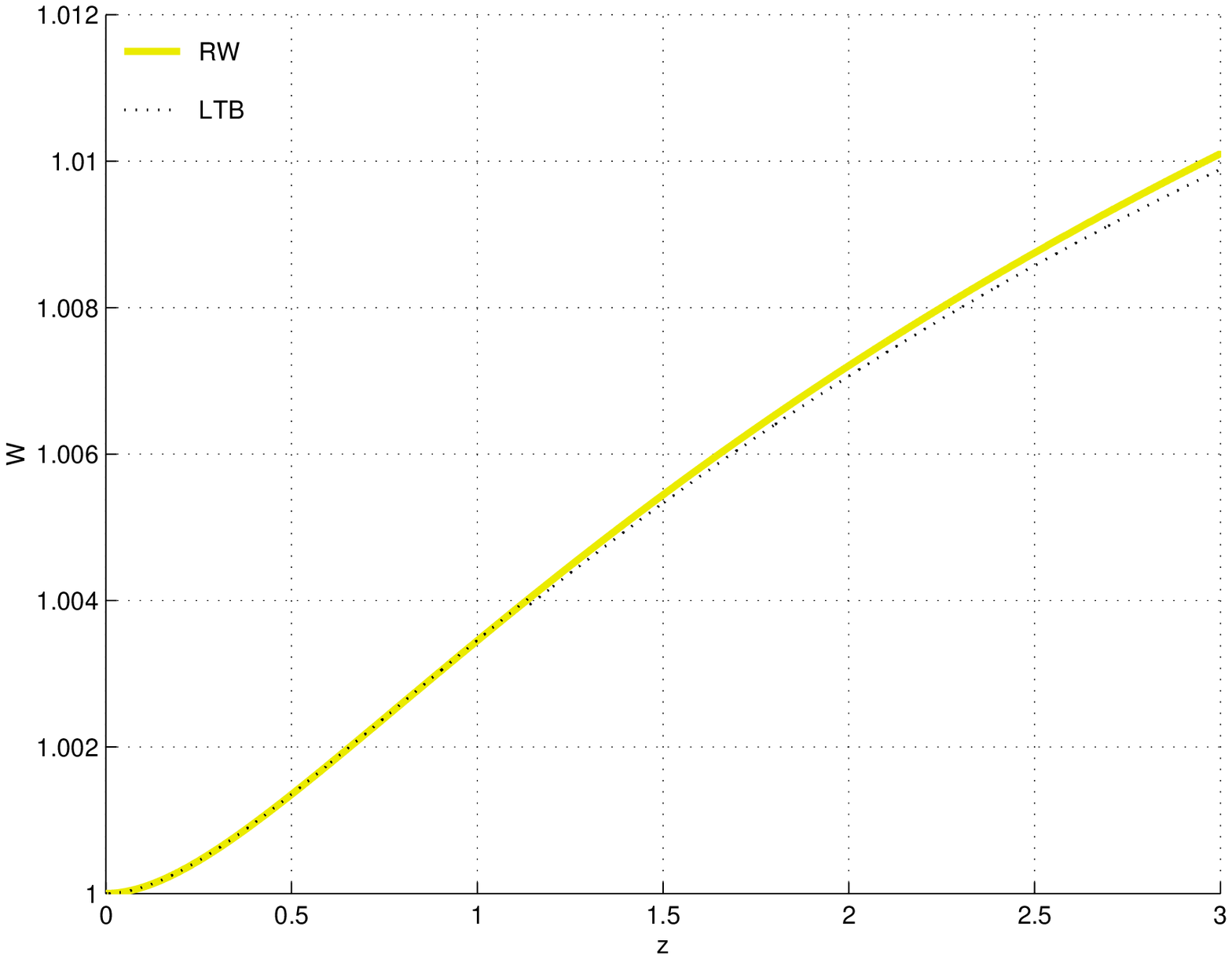}
\caption{Results of $W$ vs. $z$ with $H_{0} \approx 0.71999$, $q_{0}
\approx 0.490004$ and $\delta z = 0.001$. Matching the $M$ values at
the first connection point and the $W$ values at the second
connection point. The solid curve is the correct RW expression and
the dotted one is our numerical output using Runge-Kutta as the
integration method.}
\centering \label{home1_W_RK}
\end{figure}

Below we present the results of the comparison for a homogeneous
model with $q_{0d} = 0.49$ and $H_{0d} = 0.72$. This is the case
when we have a negatively curved universe, but very close to the
flat case. In Figures \ref{home1_r}-\ref{home1_M}, the curves
plotted from our numerical output using the Runge-Kutta method and
the ones from the generated RW data are in very good agreement with
each other, and there is only 0.02106 \% difference between the
curves in Figure \ref{home1_W_RK}. As can be seen in Figure
\ref{home1_W_RK}, there is still a jump in $W$ at $z_a$ although it
is barely visible, while the jump at the second matching point is no
longer visible. This justifies the earlier decisions to match first
$M$ and then $W$ at the two connections between the numerical
integration parts and the series expansion part.

\begin{figure}[!b]
\includegraphics[width=10.4cm, height=7.65cm]{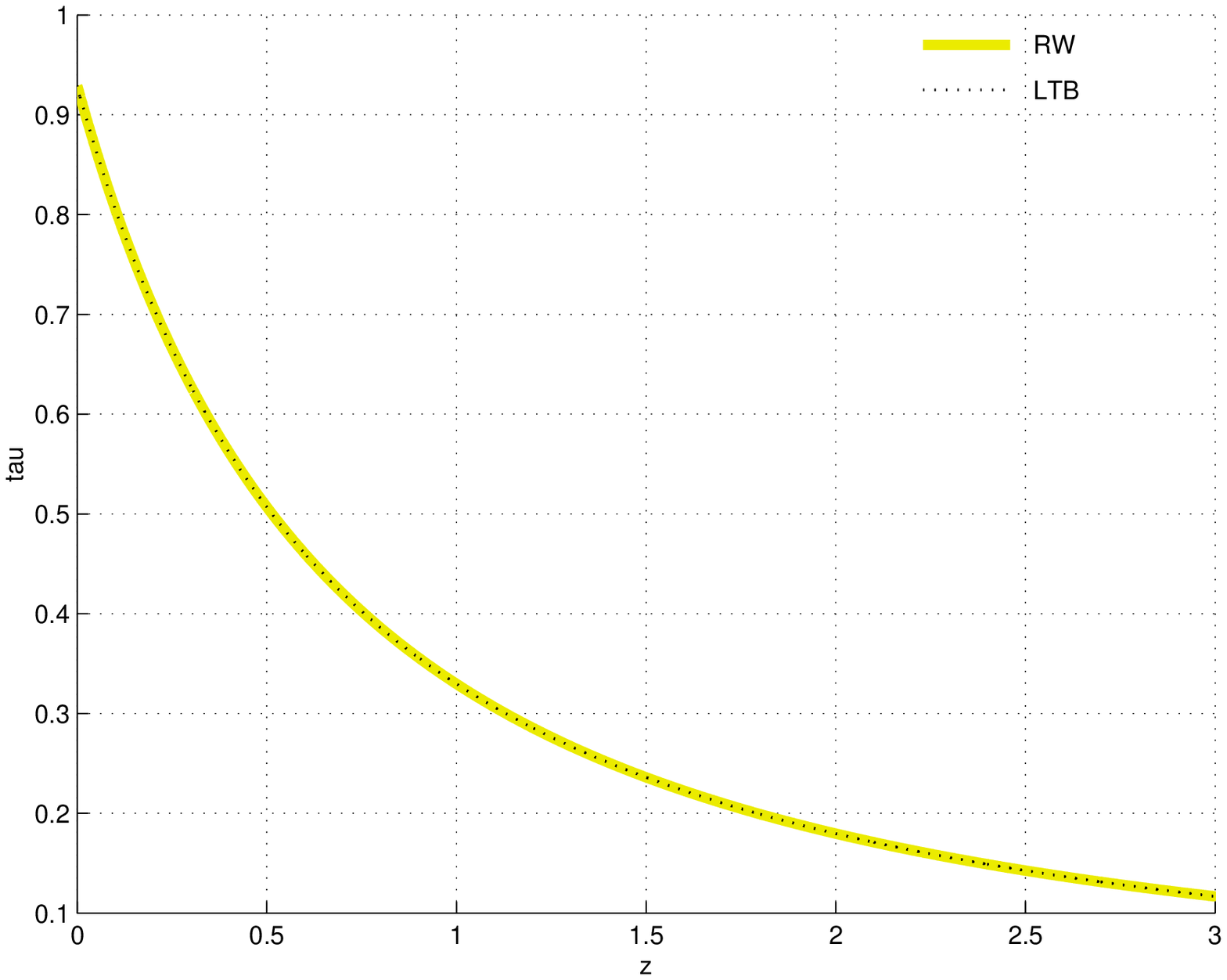}
\caption{Results of $\tau$ vs. $z$ with $H_{0} \approx 0.71999$,
$q_{0} \approx 0.490004$ and $\delta z = 0.001$. The solid grey
curve is the correct RW expression and the dotted black one is our
numerical output using Runge-Kutta as the integration method.}
\centering \label{home1_tau}
\end{figure}

\begin{figure}[!t]
\includegraphics[width=10.4cm, height=7.65cm]{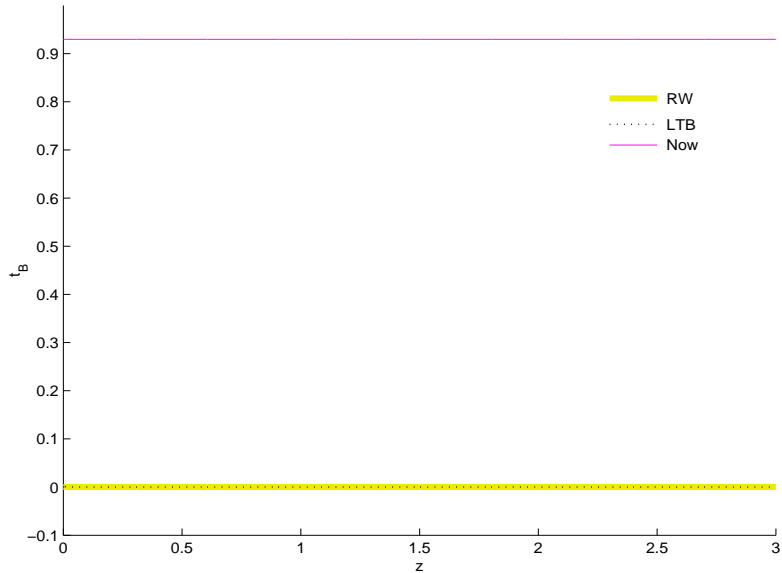}
\caption{Results of $t_B$ vs. $z$ with with $H_{0} \approx 0.71999$,
$q_{0} \approx 0.490004$ and $\delta z = 0.001$. The thick solid
grey curve is the correct RW expression and the dotted black one is
our numerical output using Runge-Kutta as the integration method.
The solid black line on the top is the current age of the universe.}
\centering \label{home1_tB}
\end{figure}

The $\tau$ curves in Figure \ref{home1_tau}, and the $t_B$ in Figure
\ref{home1_tB}, show good agreement between the numerical output and the
correct values.


\clearpage


\subsection{Inhomogeneous models}

A complete set of comparison plots from one of the inhomogeneous models tested
is given here --- a model with varying
geometry/energy. This model is one in which the
two arbitrary functions $M$ and $t_B$ take a RW form, while we vary
the third function $E$.

\begin{figure}[h]
\includegraphics[width=10.4cm, height=7.65cm]{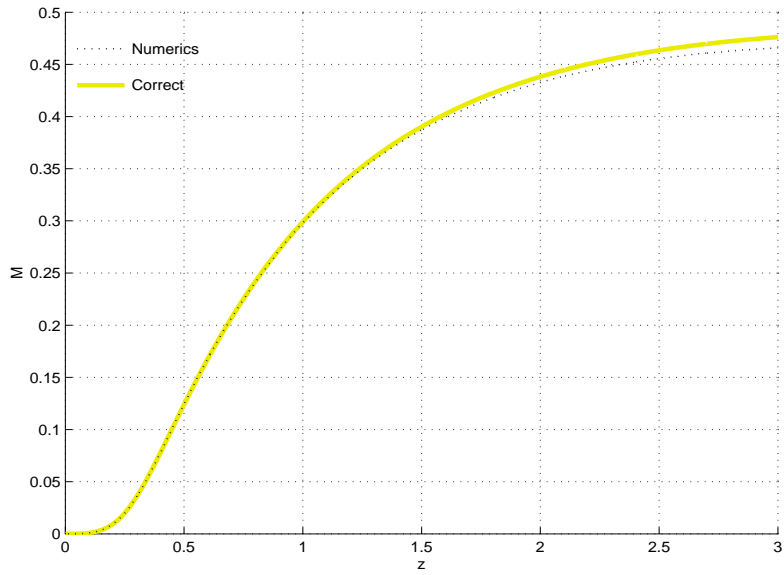}
\caption{Results of $M$ vs. $z$ with $H_{0} \approx 0.71953$, $q_{0}
\approx 0.52421$ and $\delta z = 0.001$. The solid grey curve is
from the correct testing data and the dotted black curve is our
numerical output using Runge-Kutta as the integration method.}
\centering \label{Inhomo2M}
\end{figure}

\begin{figure}[h]
\includegraphics[width=10.4cm, height=7.65cm]{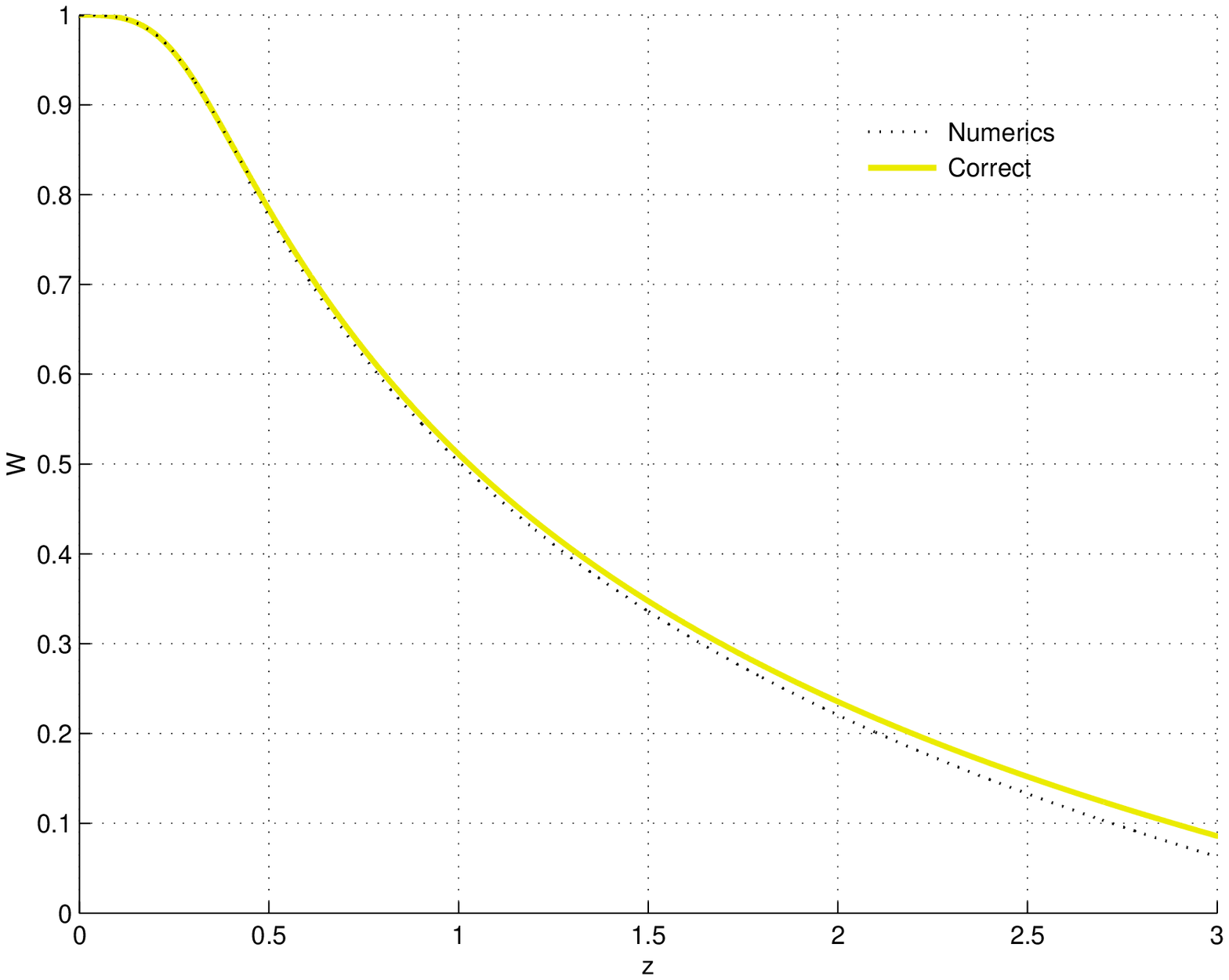}
\caption{Results of $W$ vs. $z$ with $H_{0} \approx 0.71953$, $q_{0}
\approx 0.52421$ and $\delta z = 0.001$. The solid grey curve is
from the correct testing data and the dotted black curve is our
numerical output using Runge-Kutta as the integration method.}
\centering \label{Inhomo2W}
\end{figure}

The correct origin parameters are $H_{0d} = 0.72$, $q_{0d}= 0.52$,
which gives us a near-parabolic case. The extracted values are $H_0
\approx 0.71953$ and $q_{0} \approx 0.52421$.  Figure \ref{Inhomo2M}
shows that the $M$ curve plotted from our numerical output is
slightly below the correct one; in fact there is about 2.17$\%$
error at $z = 3$. Although this percentage error is bigger than the
ones we had for the homogeneous cases, this is to be expected since
we are working with inhomogeneous data that was numerically
generated. The percentage error is a bit larger for $W$, being about
26.6$\%$ as shown in Figure \ref{Inhomo2W}. However, this percentage
error in $W$ is large mostly because $W$ is quite small, and we note
that the absolute error in $E = (W^2 - 1)/2$ is about the same as
before.

\begin{figure}[h]
\includegraphics[width=10.4cm, height=7.65cm]{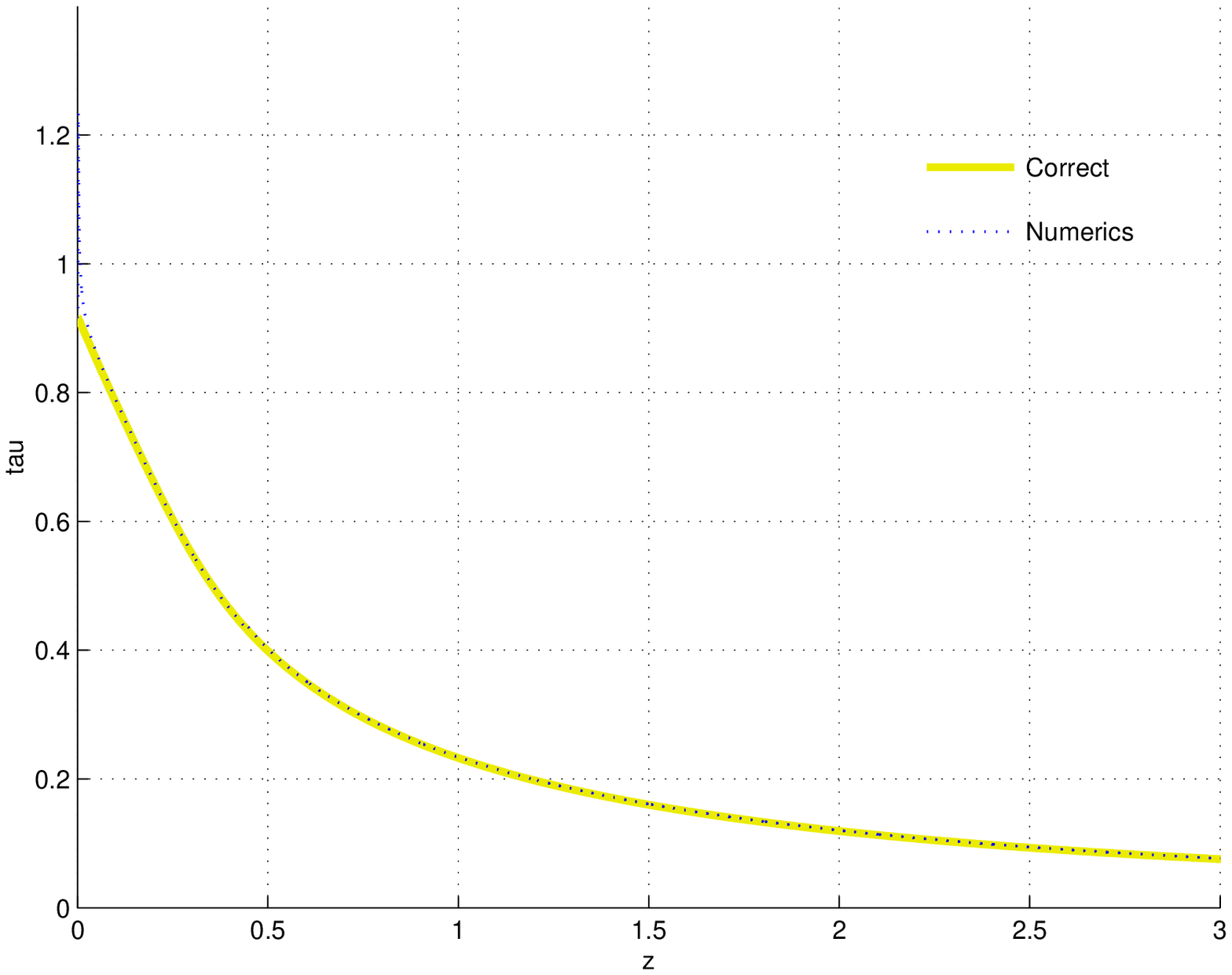}
\caption{Results of $\tau$ vs. $z$ with $H_{0} \approx 0.71953$,
$q_{0} \approx 0.52421$ and $\delta z = 0.001$. The grey curve is
from the correct testing data and the dotted black curve is our
numerical output using Runge-Kutta as the integration method.}
\centering \label{Inhomo2tau}
\end{figure}

\begin{figure}[h]
\includegraphics[width=10.4cm, height=7.65cm]{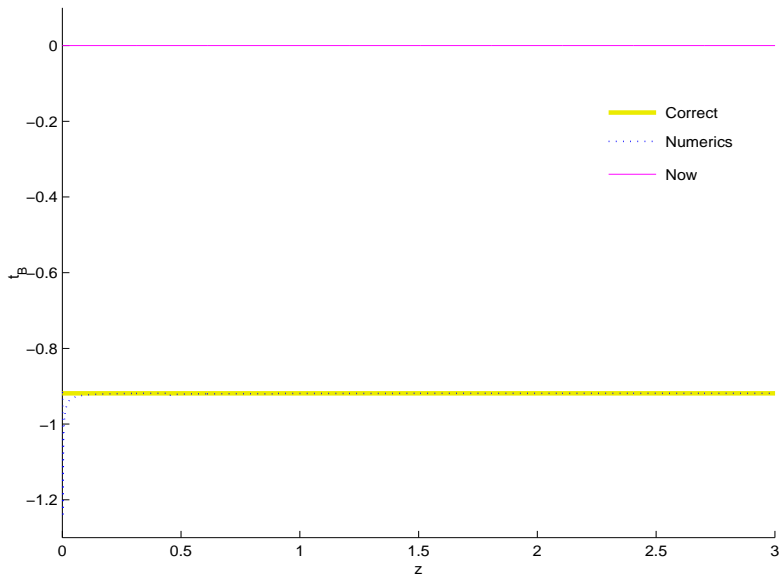}
\caption{Results of $t_B$ vs. $z$ with $H_{0} \approx 0.71953$,
$q_{0} \approx 0.52421$ and $\delta z = 0.001$. The thick solid grey
curve is from the correct correct testing data and the dotted black
one is our numerical output using Runge-Kutta as the integration
method. The solid black line is the current time (origin).}
\centering \label{Inhomo2tB}
\end{figure}

From Figures \ref{Inhomo2tau} and \ref{Inhomo2tB} we can see that
the correct data and the numerical output are generally in good agreement for
both $\tau$ and $t_B$, except near the origin. However, this is due
to insufficient accuracy in the $H_0$ and $q_0$ values deduced from the
``observational" data at $z = (3/2) \delta z$. Accurate values for $\tau$ and $t_B$
depend on an accurate $q_0$ value, which is particularly difficult
to get at the low $z$ values near the origin.  A least squares estimate
of origin values, using a wider range of near-origin data may improve accuracy here.


\clearpage

\section{Conclusion}

We have developed a computer programme to implement the MHE
algorithm. Given (spherically symmetric) data from standard
observations for redshift, apparent diameter, apparent luminosity
and galaxy number counts, as well as the associated evolution
functions, true diameter, absolute luminosity and mass per source,
it determines the metric of the (observed) universe. Its ability to
reproduce the correct metric information has been tested via
artificial data generated from both homogeneous and inhomogeneous
models.

We have started with a very simple case, in order to understand the
key elements of a numerical extraction of metric information from
observations. Obviously one does not wish to tackle the full
complexity of the problem at the start. Thus, there are still many
improvements which can be made in both the theory and the numerical
method used. Many considerations and effects must be included, for
example, source evolution theories, data set completeness, different
populations of sources, and more. At some point, a non-zero
$\Lambda$ should be considered. Also, issues like a least squares
fit for the data near the origin in order to obtain better accuracy
for $H_0$ and $q_0$, and a shorter $z$ range for the series
expansion in order to carry our numerics through the point
$\hat{R}_{max}$, also need to be dealt with for the future
development. Of course, a higher order integration method may also
be needed in the future in order to sustain the accuracy we have so
far in our numerical output out to larger $z$ values. However, any
numerical method we use to solve the DEs must be able to handle both
known data and unknown functions at a discrete set of positions.
Although higher order Runge-Kutta methods will have a natural
smoothing effect, some other form of smoothing of the data may be
needed when we tackle real data.

The bin size used for binning the observational data will affect the
accuracy of the bin averages, and require attention when one works
with the real data. Using the same bin size for the whole redshift
range, will leave the higher $z$ bins flooded with data, and low $z$
bins with very sparse data. On the other hand, making nearby
redshift bins too large may impose too much smoothing. A question
for the future then is the optimum binning strategy, and the choice
of binning versus smoothing, given that numerical integration is
ultimately a discrete process.

Our initial attempt at a numerical implementation of the MHE
procedure has successfully demonstrated the viability of the basic
concept, and opened the way to developing a more general treatment.
Our current focus is on developing a workable numerical scheme.
There are of course many relevant observational issues, such as
luminosity functions, K corrections, different source populations,
source evolution, bias, etc that must be incorporated in reducing
the observations to the data that such a programme must use. These
will be considered in the future.

Current redshift surveys do not have the accuracy or completeness to enable
meaningful metric data to be extracted at present%
\footnote{For example, the 2dF data has large fluctuations in its
number counts plot, since its thin slices were strongly affected by
the individual clusters and voids encountered. Such large
fluctuations would not be present in data averaged over all angles.
Excessive fluctuations may cause the numerical method, which
currently does not allow for angular variation, to break down,
because we take first and second differences from the $\hat{R}(z)$
data. Small fluctuations in $\hat{R}$ will generate much exaggerated
fluctuations in $d \hat{R}/dz$ and $d^2 \hat{R}/dz^2$. }.
However, the next generation of surveys is expected to provide
considerable improvements in accuracy and completeness, as well as
extending to much deeper $z$ values. Type Ia supernova measurements
hold the promise of very good luminosity distance data in the near
future, so the accuracy of luminosity functions and therefore number
counts will be the limiting factor.

Current work involves analysing the stability of the DEs, ensuring
the procedure can handle data with statistical scatter, estimating
uncertainties in the output from uncertainties in the observational
data, and using the properties of the maximum in the area distance
as a means to check or correct the result.

Eventually, knowing the metric nearby will assist in analysing more
distant observations in more than just a statistical sense, since
the spacetime that the light rays we observe have travelled through,
changes the size, brightness, frequency, position and shape of the
images we measure. Therefore, as we probe deeper into space, the
knowledge of the geometry of the universe around us will certainly
play a crucial role in the data reduction of any survey in the
future. With more reliable observational data, one may hope to
achieve one of the long term objectives of the current project --
being able to \it{prove} \rm the homogeneity of the observable
region of the universe rather than just assuming it in principle.

\ack

THCL thanks the University of Cape Town and the National Research
Foundation for their financial support.  CH thanks the National
Research Foundation for a research grant.

\appendix

\section{The near-Parabolic case}
\label{NearPara}

One may have noticed from the three evolution equations
(\ref{hyperbolic})-(\ref{elliptic}) that the parabolic evolution is
actually the $E \rightarrow 0$ limit of the other two evolutions,
which is obtained by writing the functions of $\eta$ as Taylor
expansions for small $\eta$ and noting that $\eta/\sqrt{\ E\ }$
remains finite. One can see that as we are approaching this
borderline case, the evolution equations
(\ref{hyperbolic})-(\ref{elliptic}) are not well-behaved
numerically. Also, in reality, it is very difficult, if not
impossible, to obtain an exactly parabolic case numerically. Hence,
a series expansion is needed in order to have reasonable numerical
results for the near-Parabolic case.

Most of the series expansions for the near-Parabolic case can be
found in \cite{HA05}. However, here we will consider the hyperbolic
case and give a detailed derivation following the approach in
\cite{HA05}, but for obtaining the series expansion for $\tau = t -
t_B$ only since this is the only one that is essential to us. Let us
first introduce two new variables $x \equiv 2 E/M^{2/3}$, and $a
\equiv R/M^{1/3}$. The parabolic limit now occurs when $x
\rightarrow 0$, while $R$ and $\tau$ remain finite. By (\ref{hyperbolic}),
this requires
\begin{equation}
\eta \rightarrow 0\ \mbox{and}\ \frac{\eta}{\ \sqrt{x}\ }
\rightarrow e
\end{equation}
so that the new evolution parameter $e$ remains finite for finite
$\tau$. Taylor series expansion expressions of $\tau$ and $a$ for
the hyperbolic case using equation (\ref{hyperbolic}) are just
\begin{equation}
\tau \approx \frac{\ e^3\ }{6} + \frac{\ xe^5\ }{120} +
\frac{x^2e^7}{\ 5040\ } + \frac{x^3e^9}{\ 362880\ } + \cdots
\label{tauparabolic}
\end{equation}
\begin{equation}
a \approx \frac{\ e^2\ }{2} + \frac{\ xe^4\ }{24} + \frac{\ x^2e^6\
}{720} + \frac{x^3e^8}{\ 40320\ } + \cdots \label{aparabolic}
\end{equation}
If we invert the series for $a$ by writing $e$ in series expansion
form:
\begin{equation}
e \approx e_0 + e_1x + e_2x^2 + e_3x^3 + \cdots\ , \label{eseries}
\end{equation}
then substituting into (\ref{aparabolic}), and solving for the
coefficients $e_i$, we get
\begin{equation}
\hspace{-15 mm} e \approx \sqrt{2a} \left ( 1 - \frac{1}{\ 12\ }\ ax
+ \frac{3}{\ 160\ }\ a^2x^2 - \frac{5}{\ 896\ }\ a^3x^3 +
\frac{35}{\ 18432\ }\ a^4x^4 + \cdots \right )\ , \label{eparabolic}
\end{equation}
which we substitute into (\ref{tauparabolic}), and write it in terms
of $R$, $M$ and $E$, giving
\[
\hspace{-5 mm} \tau \approx \sqrt{\frac{\ 2R^3\ }{M}} \left (
\frac{1}{\ 3\ } - \frac{1}{\ 20\ }\ \frac{\ R(2E)\ }{M} + \frac{3}{\
224\ }\ \frac{\ R^2 (2E)^2\ }{M^2} - \frac{5}{\ 1152\ }\ \frac{\ R^3
(2E)^3\ }{M^3} \right.
\]
\begin{equation}
\left. + \frac{35}{\ 22528\ }\ \frac{\ R^4 (2E)^4\ }{M^4} + \cdots
\right )\ . \label{tauhyperbolicnearparabolic}
\end{equation}
Equation (\ref{tauhyperbolicnearparabolic}) is the $\tau$ series
expansion expression for the near-parabolic case. One can do the
derivation using the elliptic evolution equations similarly.

\section{Coefficients of the series expansions near $\hat{R}_{max}$}
\label{RmaxSeries}

Let us say that $\hat{R}_{max}$ occurs at $z_{m}$, $\hat{\mu}n(z_m) =
(\hat{\mu}n)_0$, and we define $\Delta z = z - z_{m}$. So the series
expansions for $\hat{\mu}n(z)$, $\hat{R}(z)$, $\phi(z)$, $M(z)$ and
$W(z)$ have the form
\begin{equation}
\hspace{-10 mm} \hat{\mu}n(z) = (\hat{\mu}n)_{0} + (\hat{\mu}n)_{1}\Delta z +
(\hat{\mu}n)_{2}\Delta z^{2} + (\hat{\mu}n)_{3}\Delta z^{3} + \cdots\ ,
\label{Seriesmn}
\end{equation}
\begin{equation}
\hspace{-10 mm} \hat{R}(z) = \hat{R}_{max} + R_{2}\Delta z^{2} +
R_{3}\Delta z^{3} + R_{4}\Delta z^{4} + R_{5}\Delta z^{5} + \cdots\ ,
\label{SeriesRhat}
\end{equation}
\begin{equation}
\hspace{-10 mm} \phi(z) = \phi_{0} + \phi_{1}\Delta z + \phi_{2}\Delta z^{2} +
\phi_{3}\Delta z^{3} + \cdots\ , \label{SeriesPhi}
\end{equation}
\begin{equation}
\hspace{-10 mm} M(z) = M_{0} + M_{1}\Delta z + M_{2}\Delta z^{2} + M_{3}\Delta z^{3} +
\cdots\ , \label{SeriesM}
\end{equation}
and
\begin{equation}
\hspace{-10 mm} W(z) = W_{0} + W_{1}\Delta z + W_{2}\Delta z^{2} + \cdots\ .
\label{SeriesW}
\end{equation}
Hence,
\begin{equation}
\hspace{-10 mm} \frac{\ d\hat{R}\ }{dz} = 2R_{2}\Delta z + 3R_{3}\Delta z^{2} +
4R_{4}\Delta z^{3} + 5R_{5}\Delta z^{4} + \cdots\ , \label{SeriesdR}
\end{equation}
and
\begin{equation}
\hspace{-10 mm} \frac{\ d^{2}\hat{R}\ }{dz^{2}} = 2R_{2} + 6R_{3}\Delta z
+ 12R_{4}\Delta z^{2} + 20R_{5}\Delta z^{3} + \cdots\ . \label{Seriesd2R}
\end{equation}

And the expressions for the coefficients of the series expansion for
$M(z)$, $\phi(z)$ and $W(z)$ are given below:
\begin{equation}
\hspace{-10 mm} M_{0} = \frac{\ \hat{R}_{max}\ }{2}\ , \label{M0}
\end{equation}
\begin{equation}
\hspace{-10 mm} M_{1} = M_{1}\ , \label{M1}
\end{equation}
\begin{equation}
\hspace{-10 mm} M_{2} = \left ( \frac{1}{\ 1 + z_{m}\ } + \frac{\
(\hat{\mu}n)_{1}\ }{(\hat{\mu}n)_{0}} \right )\frac{\ M_{1}\ }{2} -
\frac{\ R_{2}\ }{2} - \frac{\ 8\pi^{2} (\hat{\mu}n)_{0}^{\ 2}\
}{\hat{R}_{max}}\ , \label{M2}
\end{equation}
\begin{eqnarray}
\hspace{-10 mm} M_{3} = \left ( -\frac{R_{2}}{\ 3\hat{R}_{max}\ } +
\frac{(\hat{\mu}n)_{2}}{\ 3(\hat{\mu}n)_{0}\ } + \frac{(\hat{\mu}n)_{1}}{\
3(\hat{\mu}n)_{0} \left ( 1 + z_{m} \right )\ } \right )M_{1}
\nonumber \\ - \frac{1}{\ 4\ } \left ( \frac{R_{2}}{\ 1 + z_{m}\ } +
\frac{\ (\hat{\mu}n)_{1}R_{2}\ }{(\hat{\mu}n)_{0}} + R_{3} \right ) -
\frac{\ 8\pi^{2} (\hat{\mu}n)_{0} (\hat{\mu}n)_{1}\ }{\hat{R}_{max}}\ ;
\label{M3}
\end{eqnarray}
\begin{equation}
\fl \phi_{0} = \frac{\ - \hat{R}_{max} R_{2}\ }{2\pi
(\hat{\mu}n)_{0}}\ , \label{Phi0}
\end{equation}
\begin{equation}
\fl \phi_{1} = \left ( \frac{\ (\hat{\mu}n)_{1}R_{2}\
}{(\hat{\mu}n)_{0}} - \frac{R_{2}}{\ 1 + z_{m}\ } - 3R_{3} \right )
\frac{\hat{R}_{max}}{\ 4\pi (\hat{\mu}n)_{0}\ }\ , \label{Phi1}
\end{equation}
\begin{eqnarray}
\fl \phi_{2} & = \left ( \frac{\ 3(\hat{\mu}n)_{1}R_{3}\
}{2(\hat{\mu}n)_{0}} - \frac{\ (\hat{\mu}n)_{1}^{\ 2}R_{2}\
}{2(\hat{\mu}n)_{0}^{\ 2}} - 4R_{4} + \frac{\ 2(\hat{\mu}n)_{2}R_{2}\
}{3(\hat{\mu}n)_{0}} - \frac{2R_{2}^{\ 2}}{\ 3\hat{R}_{max}\ }
\right. \nonumber \\
\fl & \left. - \frac{3R_{3}}{\ 2 \left ( 1 + z_{m} \right )\ }
+ \frac{2(\hat{\mu}n)_{1}R_{2}}{\ 3(\hat{\mu}n)_{0} \left ( 1 +
z_{m} \right )\ } + \frac{R_{2}}{\ 2 \left ( 1 + z_{m} \right )^{2}\
} \right )\frac{\hat{R}_{max}}{\ 4\pi (\hat{\mu}n)_{0}\ }\ ,
\label{Phi2}
\end{eqnarray}
\begin{eqnarray}
\fl \phi_{3} & = \left ( - \frac{\ 3(\hat{\mu}n)_{1}^{\ 2}R_{3}\
}{4(\hat{\mu}n)_{0}^{\ 2}} + \frac{(\hat{\mu}n)_{1} R_{3}}{\
(\hat{\mu}n)_{0} \left ( 1 + z_{m} \right )\ } + \frac{\
(\hat{\mu}n)_{2} R_{3}\ }{(\hat{\mu}n)_{0}} - \frac{\ 3 R_{2} R_{3}\
}{2\hat{R}_{max}} - \frac{R_{2}}{\ 4 \left ( 1 + z_{m} \right )^{3}\
} \right. \nonumber \\
\fl & - 5R_{5} - \frac{(\hat{\mu}n)_{1} R_{2}}{\ 4(\hat{\mu}n)_{0}
\left ( 1 + z_{m} \right )^{2}\ } - \frac{5(\hat{\mu}n)_{0}^{\ 2}
R_{2}}{\ 12(\hat{\mu}n)_{0}^{\ 2}\left ( 1 + z_{m} \right )\ } +
\frac{(\hat{\mu}n)_{2} R_{2}}{\ 2(\hat{\mu}n)_{0} \left ( 1 + z_{m}
\right )\ } \nonumber \\
\fl & - \frac{R_{2}^{\ 2}}{\ 2\hat{R}_{max} \left ( 1 + z_{m} \right
)\ } - \frac{2R_{4}}{\ 1 + z_{m}\ } + \frac{\ (\hat{\mu}n)_{1}^{\ 3}
R_{2}\ }{4(\hat{\mu}n)_{0}^{\ 3}} - \frac{\ 2(\hat{\mu}n)_{1}
(\hat{\mu}n)_{2} R_{2}\
}{3(\hat{\mu}n)_{0}^{\ 2}} \nonumber \\
\fl & \left. + \frac{(\hat{\mu}n)_{1} R_{2}^{\ 2}}{\ 6\hat{R}_{max}
(\hat{\mu}n)_{0}\ } + \frac{\ 2(\hat{\mu}n)_{1} R_{4}\
}{(\hat{\mu}n)_{0}} + \frac{3R_{3}}{\ 4 \left ( 1 + z_{m} \right
)^{2}\ } + \frac{\ (\hat{\mu}n)_{3} R_{2}\ }{2(\hat{\mu}n)_{0}}
\right ) \frac{\hat{R}_{max}}{\ 4 \pi (\hat{\mu}n)_{0}\ }\ ;
\label{Phi3}
\end{eqnarray}
and
\begin{equation}
\hspace{-10 mm} W_{0} = \frac{M_{1}}{\ 4\pi (\hat{\mu}n)_{0}\ }\ ,
\label{W0}
\end{equation}
\begin{equation}
\hspace{-10 mm} W_{1} = \frac{M_{1}}{\ 4\pi (\hat{\mu}n)_{0} \left ( 1
+ z_{m} \right )\ } - \frac{R_{2}}{\ 4\pi (\hat{\mu}n)_{0}\ } -
\frac{\ 4\pi (\hat{\mu}n)_{0}\ }{\hat{R}_{max}}\ , \label{W1}
\end{equation}
\begin{eqnarray}
\hspace{-10 mm} W_{2} = - \frac{R_{2}M_{1}}{\ 4\pi (\hat{\mu}n)_{0}
\hat{R}_{max}\ } - \frac{3R_{3}}{\ 16\pi (\hat{\mu}n)_{0}\ } +
\frac{(\hat{\mu}n)_{1} R_{2}}{\ 16\pi (\hat{\mu}n)_{0}^{\ 2}\ } \nonumber \\
- \frac{3R_{2}}{\ 16\pi (\hat{\mu}n)_{0} \left ( 1 + z_{m} \right )\ }
- \frac{\ 2\pi (\hat{\mu}n)_{1}\ }{\hat{R}_{max}}\ . \label{W2}
\end{eqnarray}

\section{The near origin expressions}
\label{NearOrigin}

We do series expansion of equations (A.1) and (A.2) in Appendix A of
MHE near the origin, obtaining
\begin{equation}
\hat{R} \approx \frac{1}{\ H_0\ }\ z - \frac{\ (3 + q_0)\ }{2H_0}\
z^2 + \frac{\ 4 + q_0 + q_0^{\ 2}\ }{2H_0}\ z^3 + \cdots\ ,
\label{seriesRhatRW}
\end{equation}
and
\begin{equation}
  \fl 4 \pi \hat{\mu}n \approx \frac{\ 3q_0\ }{H_0}\ z^2 - \frac{\ 6q_0(1 +
q_0)\ }{H_0}\ z^3 + \frac{\ 3q_0(15q_0^{\ 2} + 14q_0 + 13)\ }{4H_0}\
z^4 + \cdots\ .\label{series4pmnRW}
\end{equation}

We test the accuracy of the generated $H_0$ and $q_0$ values using
$z = (3/2)\delta z$, and $\hat{R}$ and $4 \pi \hat{\mu}n$ values at this same
$z$ since this is where we have the first complete set of data
according to available observational data and the way we discretise
our DEs, and therefore, find the combination with the most
consistent accuracy for different $q_0$ values. We find that in general, using both equations
(\ref{seriesRhatRW}) and (\ref{series4pmnRW}) with the same number of
terms gives us better accuracy. We can then get expressions for
$H_0$ and $q_0$ near the origin in terms of $z$, $\hat{R}$ and $4
\pi \hat{\mu} n$ only. The results are

\begin{eqnarray}
   \fl q_0 & = - \frac{\sqrt{(4 \pi \hat{\mu}n)^{2}
   + 36 \hat{R}^{2} (2 z - 1)^2
   + 12 \hat{R} (4 \pi \hat{\mu}n) (10z - 7)}}{24z\hat{R}} \nonumber \\
   \fl &~~~~~~~~~~~~ + \frac{4 \pi \hat{\mu}n + 6 \hat{R}(1 - 2z)}{24z\hat{R}}\ ,
   \label{q0RWseries}
\end{eqnarray}
and
\begin{equation}
   H_0 = \frac{3q_0z^2(1 - 2q_0z - 2z)}{4 \pi \hat{\mu}n} \ .
   \label{H0RWseries}
\end{equation}

Now we have a way of determining the origin values for $H_0$ and
$q_0$ from the data. If we need to generate values for $r$, $\phi$,
$M$ and $W$ at $z = (1/2)\delta z$ and the origin numerically from
the RW expressions given in Appendix A in MHE, then one can perform
series expansions of them too, since the values of $z$ are small.
They are:
\begin{eqnarray}
  \fl M_{RW} & \approx \frac{q_0}{\ H_0\ } z^3 - \frac{\ 3q_0(1 + q_0)\ }{2
H_0} z^4 + \frac{\ 3q_0(3 + 2q_0 + 3q_0^{\ 2})\ }{4H_0} z^5 \nonumber \\
  \fl &~~~~~~~~~~~~ - \frac{\ q_0(28q_0^{\ 3} + 12q_0^{\ 2} +
15q_0 + 25)\ }{8H_0} z^6 + \cdots \ ; \label{RW_series_M}
\end{eqnarray}
\begin{equation}
  \fl \phi_{RW} \approx \frac{1}{\ H_0\ } - \frac{\ 2 + q_0\ }{H_0} z +
\frac{\ 3q_0^{\ 2} + 4q_0 + 6\ }{2 H_0} z^2 - \frac{\ 5q_0^{\ 3} +
6q_0^{\ 2} + 6q_0 + 8\ }{2 H_0} z^3 + \cdots \ ;
\end{equation}
\begin{eqnarray}
  \fl 2E_{RW} & \approx (1 - 2q_0) z^2 - q_0 (1 - 2q_0)z^3 + \frac{1}{4}(1 -
2q_0)(5q_0^{\ 2} - 2q_0 + 1)z^4 \nonumber \\
  \fl & ~~~~~~~~~~~~ - \frac{1}{4}(1 - 2q_0)(7q_0^{\ 3} - 4q_0^{\
2} + 1)z^5 + \cdots \ ;
\end{eqnarray}
\begin{equation}
  \fl r_{RW} \approx \frac{1}{H_0}z - \frac{\ 2 + q_0\ }{2H_0}z^2 +
\frac{\ 3q_0^{\ 2} + 4q_0 + 6\ }{6H_0}z^3 - \frac{\ 5q_0^{\ 3} +
6q_0^{\ 2} + 6q_0 + 8\ }{8H_0}z^4 + \cdots \ ;
\end{equation}
and these expressions are valid for any $q_0$. For $q_0 < 1/2$ and
$q_0 > 1/2$ respectively, the $\tau$ series expansions for the RW equations take
the form:
\begin{eqnarray}
  \fl \tau_{RW} & \approx \frac{\ \sqrt{1 - 2q_0} + q_0 \ln \left( \frac{\ 1
- \sqrt{1 - 2q_0}\ }{1 + \sqrt{1 - 2q_0}} \right)\ }{H_0(1 -
2q_0)^{3/2}} - \frac{1}{\ H_0\ }z + \frac{\ 2 + q_0\ }{2H_0} z^2 \nonumber \\
  \fl & ~~~~~~~~~~~~ - \frac{\ 3q_0^{\ 2} + 4q_0 + 6\ }{6H_0} z^3 +
\frac{\ 5q_0^{\ 3} + 6q_0^{\ 3} + 6q_0 + 8\ }{8H_0} z^4 + \cdots \ ,
\end{eqnarray}
\begin{eqnarray}
  \fl \tau_{RW} & \approx \frac{\ \sin^{-1}(\frac{\ q_0 - 1\ }{q_0})q_0 +
\frac{\ \pi\ }{2}\ q_0 - \sqrt{2q_0 - 1}\ }{H_0(2q_0 - 1)^{3/2}} -
\frac{1}{\ H_0\ }z + \frac{\ 2 + q_0\ }{2H_0} z^2 \nonumber \\
  \fl &~~~~~~~~~~~~ - \frac{\ 3q_0^{\ 2} + 4q_0 + 6\ }{6H_0} z^3 +
\frac{\ 5q_0^{\ 3} + 6q_0^{\ 2} + 6q_0 + 8\ }{8H_0} z^4 + \cdots\ .
\label{RW_series_tau}
\end{eqnarray}
%

\section*{References}

\end{document}